# IAR: An Interaction-Aware Approach to Post-Seek Recovery in Adaptive HLS Streaming

Mahdi Naseri Majd[1] and Amirfarhad Farhadi[2]

[1]Faculty of Engineering, College of Farabi, University of Tehran, Iran
mahdi.naseri.majd@ut.ac.ir
[2]School of Computer Engineering, Iran University of Science and Technology, Tehran, Iran
am_farhadi@mail.iust.ac.ir

**Abstract**

After an unbuffered seek, an adaptive bitrate (ABR) player has little forward buffer at the new position, while its controller may retain a bandwidth estimate from earlier downloads. An expensive first request can delay resumption; later requests may interrupt playback while the buffer rebuilds.

We introduce Interaction-Aware Recovery (IAR), a temporary layer around the hls.js ABR controller. After an unbuffered seek, IAR discounts the inherited estimate ($\gamma$=0.15) to set a representation ceiling. Normal adaptation continues below it until 8 s of forward buffer has been rebuilt.

In a 200-run paired study spanning four network profiles, ten seek positions, and five strategies, a conservative first request reduced resume delay by a paired median of 2.008 s relative to standard hls.js. Keeping a low ceiling through recovery reduced later interruption under stronger degradation but lowered early presented bitrate. IAR and Fixed-Low-Recovery both used L0 in this campaign, so the results did not show an additional benefit from deriving the ceiling from inherited bandwidth.

In 60 additional runs under a fixed Moderate profile, we reset the estimator to six controlled scalar priors. IAR's ceiling changed from L0 to L1 and L2 at the predicted thresholds. Resume and recovery times increased at those transitions; presented bitrate varied non-monotonically and was highest at L2. No stalls occurred. These reset-created priors do not reproduce complete, naturally evolved exponentially weighted moving average (EWMA) histories.

The results show the benefits and presented-quality costs of conservative post-seek recovery, without establishing an advantage for a history-derived ceiling over a fixed low ceiling.

## 1. Introduction

### 1.1. Background and Motivation

HTTP-based adaptive video streaming makes video segments available at several bitrates or resolutions. An adaptive bitrate (ABR) player selects among these representations as network and playback conditions change [1].

A high-bitrate representation can improve displayed quality, but it may take longer to download when network capacity is limited. If a request takes longer than the available forward buffer can cover, playback stalls. The player must therefore balance delivered quality against the time available to finish each request [1].

ABR algorithms differ in the information they use. Throughput-based approaches draw on earlier transfers [1], while buffer-based approaches respond to buffered playback time [2], [3]. Hybrid controllers use both signals [4], [5], and learning-based methods such as Pensieve derive selection policies from streaming experience [6].

During uninterrupted playback, completed downloads update the bandwidth estimate as the player builds or consumes forward buffer. Several seconds of buffered media can absorb a slow download, but a request becomes more consequential as that buffer shrinks. An unbuffered seek can remove this protection almost at once.

## 1.2. The Post-Seek Recovery Problem

When a seek lands outside the buffered ranges, the player must download media at the new position before playback can resume.

Studies of seeking have documented changes in streaming behavior and perceived quality [7]. Here, we consider how a player uses bandwidth information retained from before an unbuffered seek. That information can still be useful when network conditions are similar on either side of the seek.

The difficulty arises from the combination of state persistence and reduced playback protection. If delivery conditions have changed, the historical estimate may be optimistic for the next transfer. Even when capacity has not changed, the same representation decision can become more costly because the player no longer has the forward buffer that previously absorbed variation in fragment delivery time.

Existing ABR work treats startup and low-buffer operation differently from steady-state playback. DYNAMIC, for instance, changes how throughput- and buffer-aware selection contribute under different player states [4]. A seek may require a similar distinction because the buffer can change faster than the estimated bandwidth.

## 1.3. Research Gap and Positioning

Seek-aware research has examined interaction-dependent buffering, non-linear navigation, and how seeking affects viewing experience [8], [9], [7]. We discuss these studies in Section 2.2. They do not directly address how a controller should use bandwidth information inherited across an unbuffered seek.

This study does not attempt to predict a seek destination, prefetch likely future content, or redesign the player's steady-state ABR policy. Instead, it asks whether bandwidth state inherited from before an unbuffered seek should retain the same influence immediately afterward, when the forward-buffer protection associated with the old playback position has disappeared.

We study this question in hls.js 1.7.1. Its ABR controller uses bandwidth estimates from earlier transfers together with loading and starvation considerations when selecting a level [10]. After an unbuffered seek, that information may still be available even though the forward buffer at the new position has been lost.

We do not assume that bandwidth information from before a seek is always inaccurate. Instead, we examine whether it should have less influence immediately after an unbuffered seek, while the player rebuilds forward-buffer protection at the new position.

## 1.4. Proposed Approach

We address this question with a temporary constraint on hls.js that leaves its steady-state ABR algorithm unchanged. IAR activates when a seek destination falls outside the ranges buffered immediately before the seek.

At recovery entry, IAR reads the bandwidth estimate already maintained by hls.js and discounts it using

$$B_r = \gamma B_h$$

where $B_h$ is the inherited estimate and $\gamma$=0.15 in the evaluated implementation.

IAR uses the recovery budget to set a temporary ceiling on the HTTP Live Streaming (HLS) representation ladder. It chooses the highest representation whose playlist `BANDWIDTH` value does not exceed that budget. The ceiling does not force a particular quality level: hls.js may still select any lower representation.

IAR calculates the ceiling once at recovery entry. Subsequent fragment downloads update hls.js and rebuild the forward buffer, but the ceiling remains unchanged for that recovery epoch.

Once the forward buffer reaches 8 s, IAR restores the auto-level cap that was present before recovery.

To distinguish the decisions that matter during recovery, we compare IAR with three targeted interventions and with unmodified hls.js. Lowest-on-Seek constrains only the first post-seek request. Fixed-Low-Recovery holds the minimum-level ceiling until the buffer recovers. Estimator-Reset removes the inherited bandwidth estimate without applying an IAR ceiling. Together, these comparisons separate the effects of the first request, continued recovery, and the inherited estimate.

The primary campaign includes 200 paired runs across four network profiles and ten seek positions. A separate 60-run experiment varies controlled scalar bandwidth priors while keeping the network profile fixed. That experiment tests sensitivity to the value presented at recovery entry, not to the full EWMA history of a naturally evolved estimate.

We address five questions about the initial request, the duration of recovery, the use of historical bandwidth, presented-quality costs, and sensitivity to the inherited estimate:

RQ1. How much of post-seek resume delay is determined by the representation selected for the first request after an unbuffered seek?

RQ2. Is constraining only the first post-seek request sufficient, or should a conservative restriction remain active while forward buffer is rebuilt?

RQ3. Does deriving the recovery ceiling from inherited bandwidth state provide behavior different from using a fixed conservative ceiling or resetting the estimator?

RQ4. What quality cost is introduced by conservative post-seek recovery, and how does the player return toward its pre-seek representation afterward?

RQ5. How does IAR behavior change when the inherited bandwidth-estimator state changes while the post-seek network trajectory is held fixed?

## 1.5. Main Contributions

The study contributes the following:

1. Characterization of post-seek state asymmetry. We identify an unbuffered post-seek condition in which useful forward-buffer protection at the destination is lost while network state learned before the interaction may remain available to the ABR controller.
2. Interaction-Aware Recovery. We introduce IAR, a temporary recovery layer that discounts inherited bandwidth state, converts the resulting budget into a representation ceiling, and retains that ceiling until the player has rebuilt 8 s of forward buffer without replacing the underlying steady-state ABR algorithm.
3. Mechanism-decomposition evaluation. Using Baseline, Lowest-on-Seek, Fixed-Low-Recovery, Estimator-Reset, and IAR, the primary evaluation separates the effect of the initial post-seek request from the effect of maintaining a recovery state and from the use of inherited bandwidth information.
4. Controlled historical-state sensitivity analysis. A separate experiment reinitializes the hls.js bandwidth estimator to predefined scalar priors while keeping the post-seek network trajectory fixed. This shows that IAR is not simply a fixed-low strategy: its temporary ceiling changes at the representation thresholds predicted by the controlled prior, while the experiment is explicitly not interpreted as a reproduction of complete natural EWMA histories.

Our results favor a temporary response to unbuffered seeking, with measurable costs as well as benefits. The discount and exit threshold remain specific to the player and conditions evaluated here.

# 2. Background, Related Work, and Problem Formulation

## 2.1. Adaptive Bitrate Streaming and Recovery State

An HLS master playlist describes the available representations, from which an ABR controller selects during playback [11]. These decisions trade picture quality against delivery time and rebuffering risk [1]. BBA uses buffer occupancy to guide level selection [2], while BOLA formulates a buffer-aware quality–rebuffering trade-off [3]. Measurements of deployed players have shown substantial differences in how algorithms react to network changes [12], including unstable bitrate choices under competing traffic [13].

Other work addresses different sources of instability. FESTIVE considers request scheduling and representation switching [14]. MPC combines predicted throughput with buffer occupancy [5], and Pensieve learns a policy from prior streaming experience [6]. DYNAMIC changes the roles of throughput-based control and BOLA under low-buffer conditions [4]. Throughput estimation has also received considerable attention: PANDA probes available capacity rather than using observed transfer throughput directly [15], CS2P predicts throughput using data-driven models [16], and Oboe adapts ABR parameters to network conditions [17]. These approaches help manage adaptation during playback, but none directly examines the inherited bandwidth estimate at the moment an unbuffered seek removes forward-buffer protection.

## 2.2. Seek-Aware Adaptation and Positioning of IAR

Seek-aware research has explored other consequences of viewer interaction. BufTune adjusts buffering to reduce wasted prefetching when viewers seek or skip [8]. JumpDASH combines content awareness, buffer management, and prefetching to support non-linear viewing [9]. Subjective evidence also associates seeking, rebuffering, and quality changes with perceived experience [7].

DYNAMIC already recognizes seeking and low-buffer playback as distinct operating conditions [4]. It still makes use of the available throughput estimate. We examine whether the influence of an inherited estimate should change until useful forward buffer has been rebuilt.

IAR addresses this narrower question through a temporary bandwidth-derived ceiling without predicting seek destinations or redesigning steady-state adaptation. Table 1 places this mechanism alongside buffer-based selection [3], interaction-aware buffering [8], and non-linear navigation [9].

Table 1. Positioning of IAR relative to representative prior work.

| Approach | Main signal / mechanism | Seek / interaction handling | Primary goal | Treatment of inherited pre-seek bandwidth state |
| --- | --- | --- | --- | --- |
| BOLA [3] | Buffer occupancy | Not seek-specific | Quality–rebuffering trade-off | Not explicitly addressed |
| DYNAMIC [4] | Throughput + buffer / BOLA | Includes startup, seek, and low-buffer behavior | Responsive and stable adaptation | Uses available throughput estimate during recovery |
| BufTune [8] | Interaction history + target-buffer tuning | Yes | Reduce wasted prefetched traffic | Not the main focus |
| JumpDASH [9] | Content awareness + prefetching + ABR | Yes | Support non-linear viewing / jumping | Not the main focus |
| IAR | Inherited estimator state + unbuffered-seek state + forward buffer | Yes, after an unbuffered seek | Post-seek recovery | Temporarily reduces confidence by discounting inherited state |

## 2.3. Post-Seek State Asymmetry

During sequential playback, recent transfers inform ABR decisions while buffered media provides time to tolerate variability in subsequent downloads [1].

An unbuffered seek changes this balance. Let $t_s$ denote the seek event and $B_f(t)$ the useful forward buffer at the active playback position. Immediately after a seek to an unbuffered destination,

$$B_f(t_s^+) \approx 0 \quad (1)$$

Equation (1) concerns the forward buffer at the destination. It does not require the player's entire media buffer to be empty; data retained around the previous playhead may still exist but cannot immediately support playback from the new position.

The bandwidth estimator can evolve differently. Let $\hat{C}(t)$ denote the bandwidth estimate available to the ABR controller. Before enough new post-seek transfer information has arrived, the controller may still begin recovery with approximately the state inherited from before the interaction:

$$\hat{C}(t_s^+) \approx \hat{C}(t_s^-) \quad (2)$$

Equation (2) describes the estimate available at recovery entry. The estimator can update after post-seek transfers complete.

An inherited estimate is useful when it continues to describe the link. The difficult case is a decrease in post-seek capacity, $C_{post}$, relative to the estimate:

$$C_{\text{post}} \ll \hat{C}(t_s^-) \quad (3)$$

Equation (3) identifies the capacity-mismatch case examined here; it is not a condition assumed for every seek.

The practical cost of such a mismatch depends on the representation requested and the amount of playback protection available. Let $S_{R,k}$ denote the size in bytes of fragment $k$ at representation $R$, and let the time to first byte (TTFB) for fragment k, denoted $TTFB_k$, correspond to the loader's response-start latency for that request. A first-order approximation of total fragment delivery time is

$$T_d(R,k) \approx \mathrm{TTFB}_k + \frac{8S_{R,k}}{C_{\mathrm{post}}} \quad (4)$$

From Eq. (4), larger fragments require longer transfers at the same capacity. A forward buffer may conceal part of this delay during sequential playback, but provides much less protection immediately after the unbuffered seek. Segment size can also affect delivery at a given nominal bitrate; SARA explicitly accounts for segment sizes when estimating download time [18].

Figure 1 shows the state change at the seek: the useful forward buffer falls, the historical bandwidth state remains available, and the next download has less protection against delay.

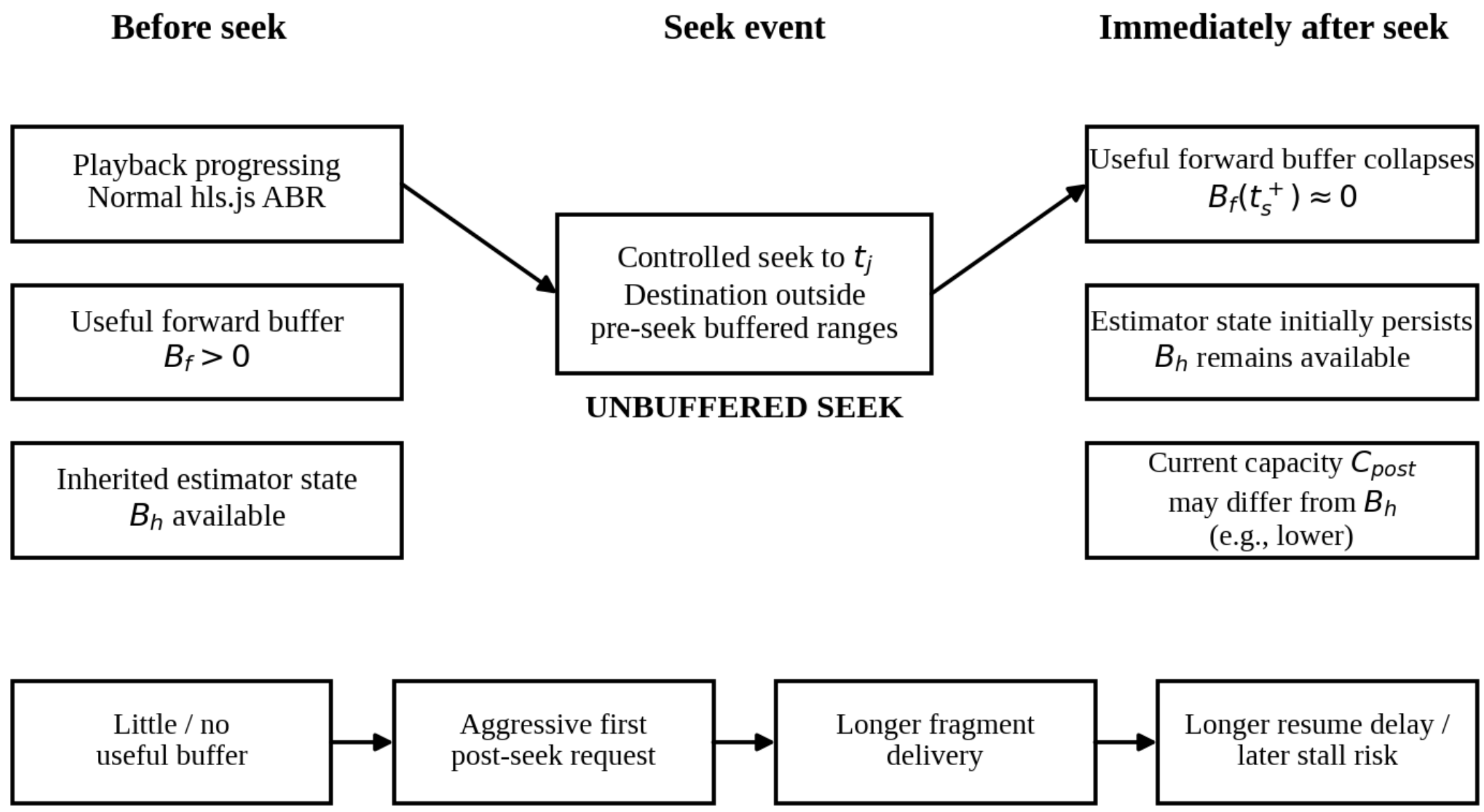

Figure 1. Post-seek state asymmetry after an unbuffered seek.

For hls.js 1.7.1, the estimate used for level selection can persist across the interaction [10]. The formal model in Eqs. (1)–(4) motivates testing a temporary restriction while playable forward buffer is rebuilt.

This problem motivates a short-lived recovery restriction on the initial request and subsequent downloads, together with measurements of its continuity and quality costs.

# 3. IAR Design

## 3.1. Design Overview

The recovery design has three parts: detecting an unbuffered seek, applying a bandwidth-derived temporary level ceiling, and releasing the ceiling once sufficient forward buffer has been rebuilt.

The hls.js ABR controller remains active throughout playback. IAR adds a temporary ceiling when a seek falls outside the ranges available immediately before the interaction, starting a new recovery epoch.

At the start of the epoch, IAR discounts the bandwidth estimate reported by hls.js and maps the reduced budget to an upper limit on the HLS representation ladder. The normal controller selects a level within that limit.

The ceiling is fixed for the epoch, even as new downloads update the hls.js estimator. IAR restores the previous auto-level cap when forward buffer at the new playback position reaches 8 s.

Figure 2 summarizes this recovery process.

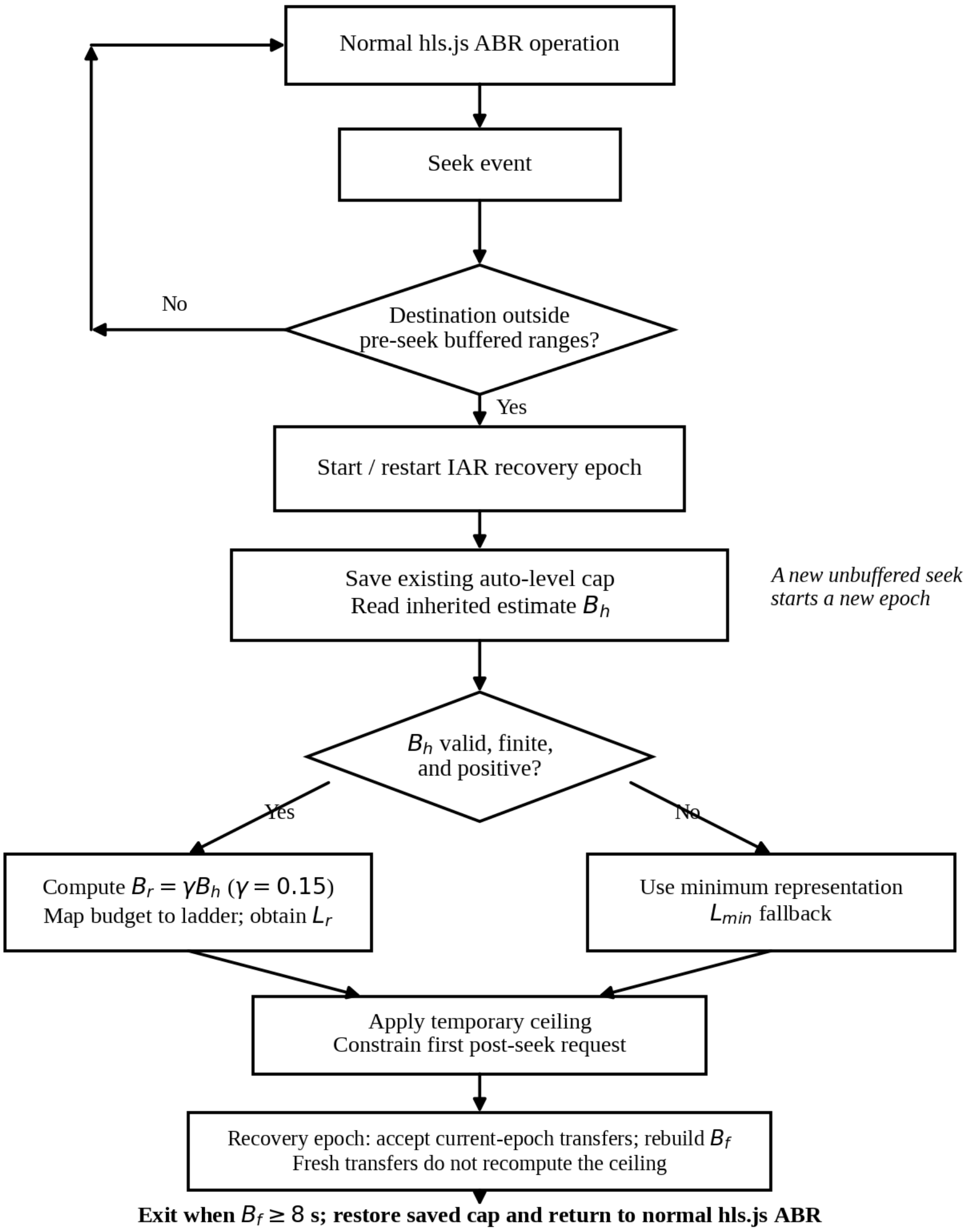

Figure 2. IAR recovery workflow and state machine.

## 3.2. Unbuffered-Seek Trigger and Recovery Epoch

IAR uses buffer availability rather than seek distance to decide when recovery is needed. A large seek may land in an already buffered region, while a shorter seek can reach data that has not been downloaded. Only the latter condition requires recovery.

Let $\mathcal{B}_{\text{pre}}$ denote the set of media positions covered by the buffered ranges immediately before the seek, and let $t_1$ denote the destination position. IAR is activated according to

$$\text{IARTrigger} = \text{SeekEvent} \wedge \left(t_1 \notin \mathcal{B}_{\text{pre}}\right) \tag{5}$$

Equation (5) activates IAR when the seek destination lies outside the buffered ranges observed before the interaction.

If the destination is already contained in $\mathcal{B}_{\text{pre}}$, the seek does not start IAR recovery and normal adaptation continues. If the destination is outside those ranges, a new recovery epoch is created.

Recovery epochs also distinguish requests from different seek regions. A transfer started before the new epoch may finish afterward, but it is not counted as a fresh post-seek transfer for the current recovery.

If another unbuffered seek occurs while recovery is already active, the existing recovery state is terminated and a new epoch is started for the new destination.

## 3.3. Historical Bandwidth Discounting

At the beginning of a recovery epoch, IAR reads the bandwidth estimate currently available to the hls.js ABR controller. We denote this inherited value by $B_h$.

Rather than using $B_h$ directly to define the temporary recovery range, IAR applies the fixed discount

$$B_r = \gamma B_h, \qquad \gamma = 0.15 \tag{6}$$

where $B_r$ is the recovery bandwidth budget.

We selected $\gamma$=0.15 during development on a separate tuning trace and fixed it before running the final primary and RQ5 campaigns.

Equation (6) discounts the inherited estimate; it is not intended to measure current post-seek capacity.

For a valid positive $B_h$, the fixed value $0<\gamma<1$ guarantees

$$B_r < B_h \tag{7}$$

Discounting the estimate does not ensure that the resulting budget is suitable for the new network conditions. IAR has no direct measurement of $C_{post}$ at recovery entry. Fragment sizes, latency, buffer availability, and the time required for playback to resume also affect the suitability of the chosen ceiling.

We therefore treat $B_r$ as a reduced historical prior, not as a new capacity estimate. RQ5 varies that prior under a fixed network profile to examine how the ceiling responds.

If the inherited bandwidth value is unavailable, non-finite, or non-positive, IAR does not cancel recovery. Instead, it falls back to the minimum representation.

## 3.4. Temporary Representation Ceiling

Let the HLS representations be ordered by the `BANDWIDTH` values declared in the master playlist:

$$P_0 < P_1 < \cdots < P_n \tag{8}$$

where $P_i$ is the BANDWIDTH value for level `i`.

IAR uses $B_r$ to select the highest representation whose playlist BANDWIDTH value does not exceed that budget. The recovery ceiling is therefore

$$L_r = \begin{cases} \max\{i \mid P_i \le B_r\} & \text{if at least one such level exists,} \\ L_{\min} & \text{otherwise} \end{cases} \tag{9}$$

If the recovery budget falls below the advertised bandwidth of the lowest representation, Eq. (9) selects that minimum level. This defines a fallback; it does not establish that the lowest representation will always download quickly enough.

During recovery, IAR restricts the selectable representation range according to

$$L_{\text{selected}} \le L_r \tag{10}$$

Equation (10) limits the highest permitted representation. Within that range, hls.js remains free to select a lower level.

The implementation applies this restriction through the hls.js auto-level capping mechanism [10]. Any pre-existing external cap is preserved: IAR does not relax a stricter restriction that was already active, and the original cap is restored when the recovery epoch ends.

IAR does not recompute $L_r$ when fresh post-seek fragments arrive. Their transfers update the underlying hls.js estimator, while the recovery constraint stays fixed until exit.

## 3.5. State-Based Recovery Exit

Releasing the ceiling after the first request can leave playback exposed before the forward buffer recovers. A fixed timeout has a different drawback: it ignores how quickly that buffer is rebuilt. IAR instead uses buffer occupancy to determine when the restriction ends.

IAR therefore uses a state-based exit rule.

Let $B_f$ denote the useful forward buffer available from the current playback position. In the final evaluated design, recovery ends when

$$B_f \geq 8\,\mathrm{s} \tag{11}$$

The recovery period consequently varies with download conditions. The cap may be released quickly when the player rebuilds 8 s of buffer, or remain longer when transfers are slower.

A tested alternative also required two fresh post-seek fragments. In the 24-run exit-rule ablation, this extra condition added no observable protection beyond the 8-s buffer criterion under the tested profiles. We therefore retained the buffer-only rule for the final campaigns.

Fresh post-seek fragment count is still recorded as a diagnostic variable and remains useful for checking recovery behavior, but it is not part of the final exit predicate.

Once Eq. (11) is satisfied, IAR removes the recovery ceiling, restores the previously saved auto-level cap, and returns unrestricted adaptation control to hls.js.

Table 2. Final IAR design settings.

| Setting | Final value / rule | Role |
|---|---|---|
| Recovery trigger | Seek destination outside pre-seek buffered ranges | Starts a new recovery epoch only when useful buffered media is unavailable at the destination |
| Historical discount γ | 0.15 | Computes $B_r=\gamma\ B_h$ |
| Representation mapping | HLS BANDWIDTH values | Maps the discounted budget to the highest permitted level |
| Below-minimum / invalid-estimate fallback | $L_{min}$ | Ensures recovery still has a defined ceiling |
| Ceiling update | Fixed within each recovery epoch | Fresh transfers do not recompute the IAR ceiling |
| Forward-buffer exit threshold | $B_f \geq 8$ s | Final recovery-exit condition |
| Fresh-fragment count $N_f$ | Diagnostic only | Retained for instrumentation and ablation analysis; not an exit requirement |

Table 2 records the settings used in the final campaigns. They were fixed for this study and should not be treated as parameters prescribed by previous ABR work.

## 3.6. Computational and Implementation Scope

IAR adds a small set of control states to an existing ABR implementation. It does not introduce a separate throughput predictor, model training, active probing, or online optimization.

The additional recovery state consists mainly of the recovery epoch, seek destination, inherited bandwidth value, recovery budget, temporary level ceiling, saved pre-recovery cap, forward-buffer state, and diagnostic information about fresh post-seek transfers.

The main operations are correspondingly simple: detect an unbuffered seek, read the existing bandwidth estimate, apply the fixed discount in Eq. (6), map the result to the representation ladder using Eq. (9), apply the temporary ceiling in Eq. (10), and monitor the buffer condition in Eq. (11).

Normal representation selection remains the responsibility of hls.js [10]. IAR only limits the maximum level that the controller can select during recovery and removes that restriction once the state-based exit condition is met.

We did not measure processor use, memory overhead, or energy consumption. Here, lightweight refers only to the limited additional control logic, not to measured computational savings.

# 4. Experimental Methodology

## 4.1. Implementation and Experimental Testbed

We implemented IAR in TypeScript using hls.js 1.7.1 [10]. The normal hls.js controller remains responsible for bitrate selection; the experiment applies the chosen seek-specific strategy when an armed unbuffered seek occurs, as described in Section 3.

The HLS presentation was hosted locally to avoid variation from public content delivery networks, server load, and Internet routing. We controlled browser-level capacity and latency with the Chrome DevTools Protocol (CDP) [19].

Playwright automated the experiments in Chromium 151.0.7922.34 [20]. We disabled browser caching so that requests experienced the configured network conditions. All final runs used the same player, media presentation, browser environment, and machine.

The test machine ran Windows 11 Home build 26200 on an 11th Gen Intel Core i5-11400H (six physical cores, 12 logical processors) with 16 GB of memory. Controlled profiles used a fixed 40-ms latency. These settings provide repeatable browser-level emulation, not a full model of a physical wired, Wi-Fi, or cellular network.

Before each run, the harness verified the hls.js version and recorded the user and effective configurations, experiment parameters, representation ladder, browser information, and event sequence. This record allows the configuration of an accepted run to be inspected. The player explicitly set only debug=false; all other hls.js settings used the effective 1.7.1 defaults logged at startup (Appendix A.1).

## 4.2. Media Presentation and Representation Ladder

All experiments used the same Big Buck Bunny source content [21]. The source was encoded as a four-level HLS presentation using H.264 video and Advanced Audio Coding (AAC) at 48 kHz stereo. Video was encoded at 30 frames/s with approximately 4-s fragments. The presentation contains 159 aligned media fragments per representation: fragments 0–157 are 4 s long and the final fragment is approximately 2.6 s, giving a total presentation duration of approximately 634.6 s.

The final master playlist advertises the following four representations. For L0, the resolution is 640×360, with `BANDWIDTH=1,178,384` bit/s and `AVERAGE-BANDWIDTH=969,405` bit/s. L1 uses 854×480, with 2,028,144 and 1,651,921 bit/s, respectively. L2 uses 1280×720, with 3,843,848 and 3,110,325 bit/s. L3 uses 1920×1080, with 7,606,104 and 6,029,471 bit/s.

The two playlist fields serve different purposes here. IAR uses BANDWIDTH to calculate the recovery ceiling in Eq. (9). Presented-quality metrics use AVERAGE-BANDWIDTH as the nominal bitrate associated with the time actually displayed at each level.

The same encoded fragments and representation ladder were used for every strategy and every network condition. The frozen run artifacts verify the codec, frame rate, audio format, aligned fragment durations, and playlist-level BANDWIDTH/AVERAGE-BANDWIDTH values reported here. They do not contain the original encoder/packager command, group of pictures (GOP) interval, container-format declaration, or independent-segment tag, so those properties are not reconstructed from memory in this manuscript; this limits bit-exact regeneration of the encoded presentation from the textual description alone.

## 4.3. Network Profiles and Paired Seek Destinations

The primary evaluation uses four controlled network profiles. Each profile begins at 9.5 Mbps. Around the controlled seek, the configured post-seek capacity is applied and acknowledged by CDP before the seek is issued. For profiles with temporary degradation, the later phase is applied approximately 20 s after the seek. Latency remains fixed at 40 ms throughout. The four profiles are summarized in Table 3.

Table 3. Primary network profiles.

| Profile | Initial capacity | Post-seek capacity | Later capacity | Purpose |
|---|---|---|---|---|
| Stable | 9.5 Mbps | 9.5 Mbps | 9.5 Mbps | Control condition with no capacity reduction |
| Moderate | 9.5 Mbps | 5.0 Mbps | 9.5 Mbps | Moderate temporary degradation |
| Severe | 9.5 Mbps | 2.0 Mbps | 9.5 Mbps | Strong temporary degradation |
| No-Recovery | 9.5 Mbps | 3.0 Mbps | 3.0 Mbps | Persistent post-seek degradation |

The primary study does not rely on a single seek location. Instead, it uses ten predefined destinations:

$$T = \{122,170,218,266,314,362,410,458,506,554\}\,\mathrm{s}$$

Each destination lies approximately 2 s after a 4-s segment boundary. The destinations are separated by 48 s, which provides spatial diversity across the presentation while retaining sufficient media after the final target for the complete observation period.

Immediately before the controlled seek, the runner verifies that the requested destination is within the media duration and outside the pre-seek buffered ranges. The seek is issued only after successful experiment arming and network-application acknowledgement. The same ten destinations are reused for all five strategies under every network profile. A pair consisting of one network profile and one seek destination therefore defines a statistical block.

The four network profiles and ten seek destinations define 40 paired blocks. Evaluating each block under all five strategies produced 200 primary runs, with every profile/seek–strategy combination represented.

## 4.4. Compared Strategies

Five strategies are included in the primary campaign.

Baseline hls.js uses the normal automatic adaptation behavior of hls.js 1.7.1. No seek-specific cap, estimator modification, or recovery state is introduced.

Lowest-on-Seek constrains only the first post-seek main-video request to L0. Immediately afterward, the explicit restriction is removed and normal hls.js adaptation continues. This strategy isolates the effect of making the first post-seek request inexpensive.

Fixed-Low-Recovery (hereafter Fixed-Low) starts the same persistent recovery state used by IAR but fixes its ceiling to the minimum representation, L0. The cap remains active until the final buffer-based recovery condition $B_f \geq 8$ s is satisfied. This strategy separates the value of maintaining a recovery phase from the value of deriving its ceiling from inherited bandwidth state.

Estimator-Reset resets the hls.js bandwidth estimator to the player's configured default estimate of 0.5 Mbps at the controlled seek and then allows standard hls.js adaptation to continue without an IAR recovery ceiling. In hls.js 1.7.1, assigning this value through bandwidthEstimate resets the underlying estimator. This provides additional context for whether discarding the inherited estimator state behaves differently from temporarily discounting it.

IAR reads the inherited bandwidth state $B_h$, computes the recovery budget $B_r$=0.15$B_h$, and derives the temporary level ceiling using Eq. (9). The first post-seek request is forced to respect that ceiling, after which normal hls.js ABR continues within the same persistent cap. The ceiling remains fixed until the forward buffer reaches the final recovery threshold. The implementation of the persistent cap and the distinction between IAR and Fixed-Low are explicit in the final player code.

The strategy-specific action is executed only after the harness has accepted the seek as the armed unbuffered experimental seek. For example, Estimator-Reset performs its reset at this point, while Lowest-on-Seek forces the next load level to L0.

The frozen primary campaign plan contains all five strategies and uses `buffer-only` recovery for both IAR and Fixed-Low-Recovery.

## 4.5. Recovery-Exit Ablation and Parameter Freeze

Before the final primary campaign, the recovery-exit rule was evaluated separately. Three candidate rules were considered: a fragment-only rule requiring at least two fresh post-seek fragments, a buffer-only rule requiring $B_f \geq 8$ s, and a combined rule requiring both conditions. The purpose of this ablation was to determine whether the fresh-fragment requirement added useful protection beyond the forward-buffer requirement.

A separate matched ablation contained 24 runs: three exit rules (fragment-only, buffer-only, and combined) evaluated at four seek targets (122, 266, 410, and 554 s) under Moderate and Severe profiles. Buffer-only and combined behaved almost identically in the tested conditions: across their eight matched runs per rule, median recovery duration was 5.342 s versus 5.328 s, median exit buffer was 8.274 s versus 8.292 s, and each rule had two stall-positive runs. Fragment-only ended recovery much earlier (median recovery 2.938 s; median exit buffer 4.605 s) and had the same stall incidence in this small ablation. This does not show that fragment-only is worse than buffer-only. Rather, it shows that adding a fragment-count requirement to the buffer criterion provided no observable extra protection in the tested conditions. We retained buffer occupancy as the final exit signal because it directly measures the playback protection that the recovery phase is intended to rebuild. A fixed fragment count is only an indirect proxy, and its relationship to available protection changes with segment duration and ongoing playback consumption. Appendix A.2 reports the quantitative ablation summary and profile-specific values.

The final campaign was executed only after this choice had been frozen. All 80 persistent-recovery runs in the primary dataset, comprising 40 IAR and 40 Fixed-Low-Recovery runs, used buffer-only, and the smallest recorded forward buffer at recovery exit was 8.013 s.

Fresh-fragment counts were retained in the logs as diagnostic information, but they were not used to determine recovery completion in the final evaluation.

## 4.6. Evaluation Metrics and Timing

Resume delay is measured from the controlled seek to the first valid post-seek frame that is actually presented and advances playback. The implementation uses `requestVideoFrameCallback()` rather than the Hypertext Markup Language (HTML) `playing` event for this boundary. If $t_{seek}$ is the monotonic time of the controlled seek and $t_{resume}$ is the time of the first valid advancing post-seek frame, resume delay is

$$T_r = t_{\text{resume}} - t_{\text{seek}} \tag{12}$$

Equation (12) is the primary measurement of how long the user-visible interruption lasts before playback begins again.

Post-resume stall duration, $T_s$, is the cumulative duration of positive stall intervals recorded after the first resumed frame and before the end of the experiment's observation horizon. A binary stall-occurrence variable is also retained.

Combined interruption is defined as

$$T_i = T_r + T_s \tag{13}$$

Equation (13) is used only as a technical continuity measure. It is not treated as a perceptual QoE score.

Presented quality is measured from media that was actually displayed, rather than from downloaded fragments alone. Two wall-clock observation windows are used: W20 covers the first 20 s after $t_{resume}$, and W40 covers the first 40 s after $t_{resume}$. At each window boundary, the endpoint corresponds to the last media timestamp actually observed through `requestVideoFrameCallback()`, rather than simply `video.currentTime`. The completed primary campaign achieved full presented-quality coverage in all audited W20 and W40 windows.

Let $d_j$ be the amount of presented media time attributed to quality segment `j`, and let $A_j$ be the HLS `AVERAGE-BANDWIDTH` of its representation. The time-weighted presented bitrate over a window `W` is

$$\overline{B}_W = \frac{\sum_{j \in W} d_j A_j}{\sum_{j \in W} d_j} \tag{14}$$

Equation (14) therefore weights quality by media that was genuinely presented to the viewer.

Quality instrumentation additionally records representation-switch count, cumulative level-step magnitude, and the fraction of presented media time spent at each representation. A quality return is confirmed only after the presented representation has reached at least the pre-seek level and sustained that condition for 4 s of presented media. Return time is measured from $t_{resume}$. Runs that do not satisfy the sustained-return condition within W40 are treated as censored rather than being assigned an artificial 40-s return time.

Recovery duration is measured from recovery entry until the $B_f \geq 8$ s exit condition is reached. Recovery-exit buffer and fresh-fragment count are retained as diagnostic measurements.

Fragment-level timing is also logged. The reported TTFB value corresponds to the hls.js loader's response-start latency, computed as `loading.first – loading.start`; transfer duration is `loading.end – loading.first`; and total request duration is `loading.end – loading.start`. These quantities are used for mechanism-level diagnostics rather than as replacements for the frame-based resume metric. Table 4 summarizes the final evaluation metrics.

Table 4. Final evaluation metrics.

| Metric | Definition | Main role |
|---|---|---|
| Resume delay $T_r$ | Seek → first valid advancing post-seek presented frame | Initial recovery |
| Post-resume stall duration $T_s$ | Sum of positive post-resume stall intervals | Continuity after resume |
| Combined interruption $T_i$ | $T_r+T_s$ | Technical interruption |
| W20/W40 presented bitrate | Time-weighted AVERAGE-BANDWIDTH of actually presented media | Quality cost |
| Switch count | Number of presented-level transitions | Adaptation stability |
| Cumulative switch magnitude | Sum of absolute level-step changes | Adaptation volatility |
| Sustained quality return | Same-or-higher pre-seek level sustained for 4 s of presented media | Quality recovery |
| Recovery duration | Recovery entry → $B_f \geq 8$ s | Duration of intervention |

## 4.7. Experimental Protocol and Controlled Historical-State Study

Each primary run begins with sequential playback under the profile's initial network condition so that the normal player and estimator state can develop before the experimental interaction. The scheduled post-seek network condition is then applied through CDP. The automation waits for successful completion of that command, records a browser-side network marker, and only then arms and issues the controlled seek. A short guard interval separates the acknowledged network transition from the seek.

The player records `FRAG_LOADING`, loader response-start timing, `FRAG_LOADED`, and the first valid resumed frame. For Stable, Moderate, and Severe, the later network phase is applied approximately 20 s after the seek. In No-Recovery, the degraded capacity remains in effect. W20 and W40 are independent quality-observation windows beginning at the first resumed presented frame, not at the network transition or seek itself.

Figure 3 summarizes the paired multi-seek protocol and event timing used in the final campaigns.

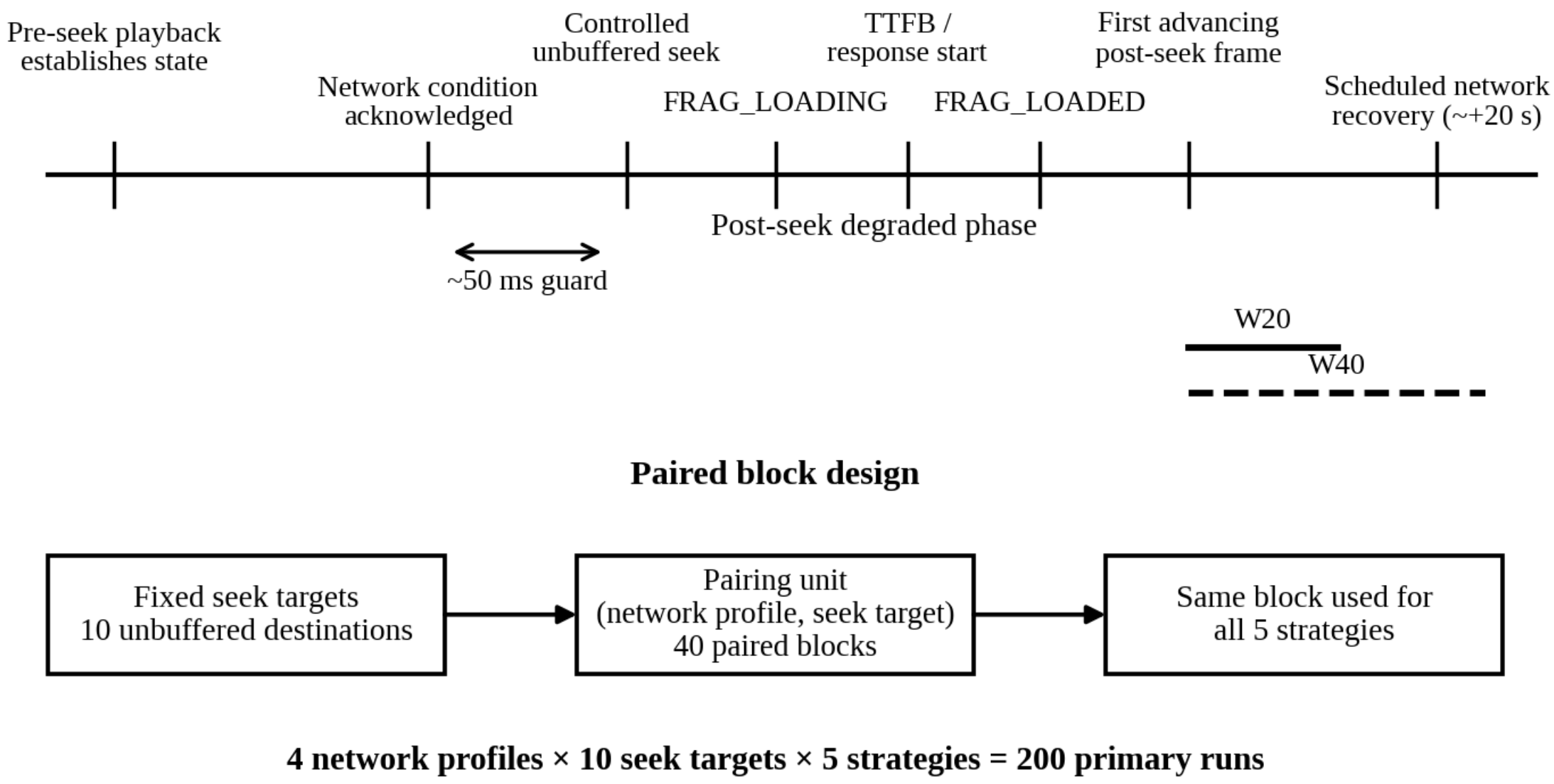

Figure 3. Paired multi-seek experimental protocol and event timing.

The campaign order was frozen before execution and interleaved conditions so that all observations from one strategy were not collected consecutively. Within a given (network profile, seek target) block, the five strategies form the paired observations used in the final statistical analysis. The completed campaign contains no missing blocks, no duplicate strategies within blocks, and no target mismatches.

A separate controlled experiment addresses RQ5. Its purpose is to determine whether IAR responds to the bandwidth value available at recovery entry rather than behaving as a fixed L0 recovery rule. The network trajectory is held fixed at Moderate, and immediately before the controlled seek the hls.js bandwidth estimator is reinitialized to a predefined scalar prior.

The injected values are

$$B_h \in \{6,10,15,20,25,30\}\ \text{Mbps}$$

With γ=0.15, these values produce recovery budgets of 0.90, 1.50, 2.25, 3.00, 3.75, and 4.50 Mbps, respectively. For the four-level BANDWIDTH ladder used here, the expected ceilings are L0, L0, L1, L1, L1, and L2.

The injection is a controlled estimator reinitialization used for causal sensitivity analysis. Assigning the predefined value through hls.bandwidthEstimate resets the hls.js estimator and creates a new estimator initialized with that scalar prior; it therefore does not reproduce the complete fast/slow EWMA history of a naturally evolved state. The runner reads back the applied value and requires positive acknowledgement before arming and issuing the seek. Each of the six priors was evaluated at all ten seek destinations, yielding 60 RQ5 runs.

All 60 completed runs contained the requested/applied estimator acknowledgement, the armed unbuffered seek, recovery start and end, first resumed frame, and complete W20/W40 instrumentation.

## 4.8. Statistical Analysis

The final experimental design is paired: the same seek destination under the same network profile is evaluated under each strategy. The inferential analysis therefore preserves this pairing instead of treating the observations as independent groups.

Continuous paired outcomes are compared using the exact Wilcoxon signed-rank distribution [22]. Zero paired differences are omitted from the signed-rank calculation. Effect size is reported using the paired rank-biserial correlation [23]. In addition to the test statistic, paired median differences are reported with exact nonparametric order-statistic confidence intervals having coverage of at least 95%.

Binary paired outcomes, including stall occurrence and return-within-W40, are evaluated using the exact McNemar test based on discordant pairs [24]. Holm adjustment [25] is applied within each predefined research-question family rather than globally across unrelated questions.

Descriptive summaries report the median and interquartile range (IQR), together with the mean ± sample standard deviation, so that the central paired pattern and raw dispersion remain visible.

The primary contrast families follow the mechanism decomposition used by the research questions. RQ1 compares Baseline with Lowest-on-Seek; RQ2 compares Lowest-on-Seek with Fixed-Low-Recovery; RQ3 compares IAR with Fixed-Low-Recovery and, separately, Estimator-Reset; and RQ4 compares presented-quality outcomes of IAR with the other primary strategies. Because runs that do not recover the pre-seek representation by W40 are right-censored, raw return delays are reported descriptively only among runs that return. Inferential testing instead uses the paired binary outcome return-within-W40. These are the methods used in the frozen primary analysis.

For RQ5, the main inferential contrasts are the estimator-state transitions that cross a representation-ceiling boundary: 10→15 Mbps for L0→L1 and 25→30 Mbps for L1→L2. States that map to the same ceiling are also inspected descriptively as plateau checks.

## 4.9. Dataset Validation, Preprocessing, and Reproducibility

Each primary run was audited before statistical analysis. Required events included the hls.js configuration snapshot, run start, initial and later network markers, successful seek arming, one controlled unbuffered seek, exactly one first post-seek fragment request, a valid first presented post-seek frame, and complete W20/W40 quality instrumentation. Strategy-specific checks verified the IAR discount, persistent-recovery mode, Estimator-Reset occurrence, Lowest-on-Seek occurrence, and buffer-based exit behavior.

The final primary audit accepted all 200 runs. No run was missing, duplicated, associated with the wrong seek target, or rejected because of a fatal hls.js error. Four additional 0.1-s playback-position changes generated internally by hls.js during stalls were recorded as unarmed seeks and correctly excluded from the experimental-seek logic.

The audit identified eight impossible negative stall durations between -0.7 and -0.5 ms. These arose because the original STALL_START timestamp used performance.now() while the corresponding end point was sampled from the requestVideoFrameCallback() callback timestamp. Because negative physical duration is impossible, the eight sub-millisecond records were clamped to 0 ms and were not classified as stalls; all other positive primary stalls were at least 55 ms. An offline perturbation sensitivity check then shifted every positive RQ2 stall duration by the worst observed clock offset (0.7 ms) in opposite directions for Lowest-on-Seek and Fixed-Low. Stall/non-stall classifications were unchanged, and the exact Wilcoxon p-values for RQ2 stall duration ($2.6703\times10^{-5}$ before Holm adjustment) and combined interruption ($9.4117\times10^{-4}$) were identical under both adverse perturbation directions. The corresponding Holm-adjusted conclusions therefore remain $p=1.07\times10^{-4}$ and $p=0.00282$. The logger was nevertheless corrected to use performance.now() consistently for STALL_END before RQ5. Appendix A.3 summarizes this sensitivity check.

The 60-run RQ5 campaign was audited independently. It contained all 60 expected block/state combinations, no failed conditions, no fatal hls.js errors, complete W20/W40 instrumentation, and exact agreement between requested and applied estimator initialization.

Campaign metadata includes the configuration and plan information required to reproduce the final runs. The resulting analysis therefore uses 200 frozen primary runs plus 60 controlled RQ5 runs. Earlier development-only runs are excluded from the final inferential evaluation.

The IAR player implementation, experiment automation scripts, frozen campaign plans, and released experimental results are available at https://github.com/mahdimnm/iar-post-seek-recovery. The repository provides instructions for setting up and running the experiments and links to the separately archived HLS media used for the reported measurements.

# 5. Results and Discussion

We report the 200-run primary campaign first, followed by the 60-run controlled historical-state study. Development runs are excluded.

The primary dataset contains 40 paired blocks, each defined by a network profile and seek position and tested with all five strategies. Tables summarize median and interquartile range (IQR) values; paired comparisons use the tests described in Section 4.8.

Table 5 provides the main continuity results from the primary campaign. All 200 runs passed the final structural audit and were retained.

Table 5. Primary continuity results.

| Profile | Strategy | Resume delay $T_r$, s, median [IQR] | Runs stalled | Stall duration $T_s$, s, median [IQR] | Combined interruption $T_i$, s, median [IQR] |
|---|---|---|---|---|---|
| Stable | Baseline | 2.644 [2.589, 2.759] | 0/10 | 0.000 [0.000, 0.000] | 2.644 [2.589, 2.759] |
| Stable | Lowest-on-Seek | 0.516 [0.490, 0.521] | 0/10 | 0.000 [0.000, 0.000] | 0.516 [0.490, 0.521] |
| Stable | Fixed-Low | 0.513 [0.490, 0.519] | 0/10 | 0.000 [0.000, 0.000] | 0.513 [0.490, 0.519] |
| Stable | Estimator-Reset | 0.520 [0.489, 0.533] | 0/10 | 0.000 [0.000, 0.000] | 0.520 [0.489, 0.533] |
| Stable | IAR | 0.510 [0.486, 0.533] | 0/10 | 0.000 [0.000, 0.000] | 0.510 [0.486, 0.533] |
| Moderate | Baseline | 2.903 [2.850, 2.929] | 0/10 | 0.000 [0.000, 0.000] | 2.903 [2.850, 2.929] |
| Moderate | Lowest-on-Seek | 0.889 [0.847, 0.910] | 1/10 | 0.000 [0.000, 0.000] | 0.889 [0.847, 0.972] |
| Moderate | Fixed-Low | 0.895 [0.840, 0.907] | 0/10 | 0.000 [0.000, 0.000] | 0.895 [0.840, 0.907] |
| Moderate | Estimator-Reset | 0.906 [0.843, 0.924] | 0/10 | 0.000 [0.000, 0.000] | 0.906 [0.843, 0.924] |
| Moderate | IAR | 0.892 [0.843, 0.915] | 0/10 | 0.000 [0.000, 0.000] | 0.892 [0.843, 0.915] |
| Severe | Baseline | 4.074 [3.968, 4.138] | 6/10 | 0.066 [0.000, 0.204] | 4.181 [4.104, 4.327] |
| Severe | Lowest-on-Seek | 2.076 [1.964, 2.124] | 6/10 | 1.540 [0.000, 1.953] | 3.626 [2.044, 4.100] |
| Severe | Fixed-Low | 2.057 [1.970, 2.121] | 6/10 | 0.059 [0.000, 0.195] | 2.175 [2.053, 2.349] |
| Severe | Estimator-Reset | 2.070 [1.965, 2.107] | 5/10 | 0.028 [0.000, 0.190] | 2.227 [1.980, 2.328] |
| Severe | IAR | 2.068 [1.959, 2.139] | 7/10 | 0.083 [0.017, 0.259] | 2.297 [2.123, 2.332] |
| No-Recovery | Baseline | 3.422 [3.352, 3.467] | 0/10 | 0.000 [0.000, 0.000] | 3.422 [3.352, 3.467] |
| No-Recovery | Lowest-on-Seek | 1.407 [1.341, 1.445] | 10/10 | 3.189 [1.524, 5.275] | 4.705 [2.899, 6.666] |
| No-Recovery | Fixed-Low | 1.407 [1.335, 1.447] | 0/10 | 0.000 [0.000, 0.000] | 1.407 [1.335, 1.447] |
| No-Recovery | Estimator-Reset | 1.429 [1.336, 1.458] | 0/10 | 0.000 [0.000, 0.000] | 1.429 [1.336, 1.458] |
| No-Recovery | IAR | 1.429 [1.354, 1.453] | 0/10 | 0.000 [0.000, 0.000] | 1.429 [1.354, 1.453] |

The values in Table 5 are computed from the frozen run-level dataset after applying the audited stall-timestamp preprocessing described in Section 4.9.

## 5.1. RQ1: Effect of the First Post-Seek Request

For RQ1, we compare standard hls.js (Baseline) with Lowest-on-Seek to isolate the first post-seek request.

Baseline requested L3 for the first post-seek fragment in all 40 blocks, whereas Lowest-on-Seek requested L0 in all 40. This difference had a large effect on resume time. The paired Baseline-minus-Lowest resume-delay difference had a median of 2008.1 ms, with a 95% interval of [2002.2, 2019.8] ms. The exact Wilcoxon signed-rank test gave a rank-biserial effect of 1.0 and a Holm-adjusted $p=7.28\times10^{-12}$.

Figure 4A shows that this approximately 2-s advantage was not confined to one network profile. The median paired resume-delay difference was about 2.156 s in Stable, 2.008 s in Moderate, 1.997 s in Severe, and 2.008 s in No-Recovery.

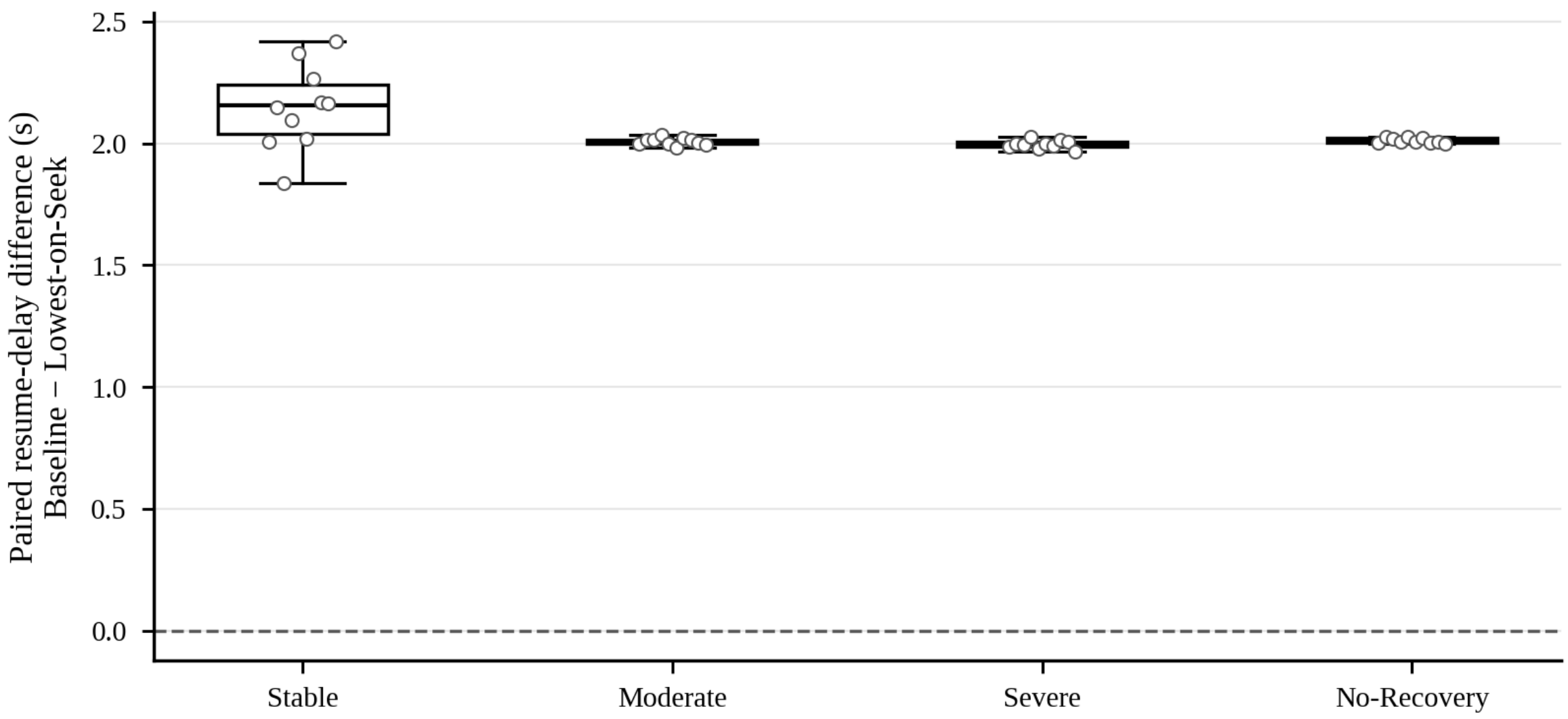

Figure 4A. First-request effect (RQ1): paired Baseline-minus-Lowest resume-delay differences across network profiles.

An important distinction appears when the requested representation is compared with the representation that actually reaches the display. Although Baseline requested L3 first in all 40 runs, its first presented post-seek level was L0 in 32 runs and L3 in only eight. This occurred because the player could abandon or replace the aggressive request before the first post-seek frame was presented. The result therefore supports the mechanism more precisely as an expensive initial request problem, rather than implying that the first displayed frame must itself be high quality.

The faster first request did not always improve continuity after resumption. Across the 40 paired blocks, four stalls occurred only under Baseline, whereas 15 occurred only under Lowest-on-Seek. The exact McNemar comparison gave $p_{Holm}=0.0192$. The pooled paired difference in combined interruption still favored Lowest-on-Seek, with a median Baseline-minus-Lowest difference of 1997.8 ms, but the effect was strongly profile-dependent.

The trade-off is clearest under No-Recovery. Baseline had no stalls in the ten blocks, while Lowest-on-Seek stalled in all ten. Median combined interruption was 3.422 s for Baseline and 4.705 s for Lowest-on-Seek. The median paired Baseline-minus-Lowest difference was approximately -1.171 s. Here, interruptions after resumption outweighed the faster initial restart with Lowest-on-Seek in several blocks.

RQ1 therefore shows that making the first post-seek request conservative strongly shortens resume delay, but resume delay alone does not determine the total continuity outcome.

## 5.2. RQ2: Does Recovery Need to Continue Beyond the First Fragment?

RQ2 compares Lowest-on-Seek with Fixed-Low-Recovery. Both begin recovery at L0; the difference is that Lowest-on-Seek removes the restriction after the first request, whereas Fixed-Low retains the L0 ceiling until $B_f\geq8$ s.

The two strategies produced essentially the same initial resume performance. The paired Lowest-minus-Fixed-Low resume-delay difference was only 1.7 ms, with a 95% interval of [-4.9, 7.6] ms and $p_{Holm}=0.4457$. This indicates that extending the restriction beyond the first fragment did not materially change the time required to restart playback.

The difference appears after playback resumes. Fourteen paired blocks contained a stall only under Lowest-on-Seek, compared with three blocks containing a stall only under Fixed-Low. The exact McNemar test gave $p_{Holm}$=0.0255. The paired stall-duration comparison also favored persistent recovery, with $p_{Holm}=1.07\times10^{-4}$.

The effect is especially visible in the stronger degradation profiles. Under Severe, both strategies recorded a positive stall in six of ten runs, but the duration of those events was markedly different: Lowest-on-Seek had a median stall duration of 1.540 s, whereas Fixed-Low had a median of only 0.059 s. The median paired difference in combined interruption was approximately 1.541 s in favor of Fixed-Low.

No-Recovery provides the clearest separation. Lowest-on-Seek stalled in all 10 runs, with a median stall duration of 3.189 s and median combined interruption of 4.705 s. Fixed-Low remained stall-free in all ten runs and had a median combined interruption of 1.407 s. The median paired Lowest-minus-Fixed-Low combined-interruption difference was approximately 3.196 s.

The profile-specific pattern is the primary interpretation of RQ2: persistence has little measurable effect under Stable and Moderate, reduces post-resume interruption under Severe, and has a pronounced benefit under persistent No-Recovery degradation. As a secondary pooled summary, the paired combined-interruption difference across all 40 blocks was 14.0 ms, with a broad 95% interval of [1.7, 1529.7] ms and $p_{Holm}$=0.00282. The width of this interval reflects the strong heterogeneity across profiles, so the pooled value is not treated as a single profile-independent effect size.

These results do not imply that persistent recovery eliminates every short interruption: brief positive stall events still occurred under Severe for Fixed-Low and IAR. Rather, the evidence shows that a first-fragment-only intervention can shift substantial interruption into the period after playback resumes, whereas maintaining the conservative ceiling while forward buffer is rebuilt reduces that risk.

The state-based nature of recovery is also visible in IAR's recovery duration. Its median duration was approximately 1.49 s in Stable, 2.63 s in Moderate, 8.29 s in Severe, and 5.58 s in No-Recovery. Thus, the recovery interval is not a fixed timeout; it adapts to how quickly the 8-s forward-buffer threshold can be rebuilt. Figure 4B shows the paired combined-interruption comparison.

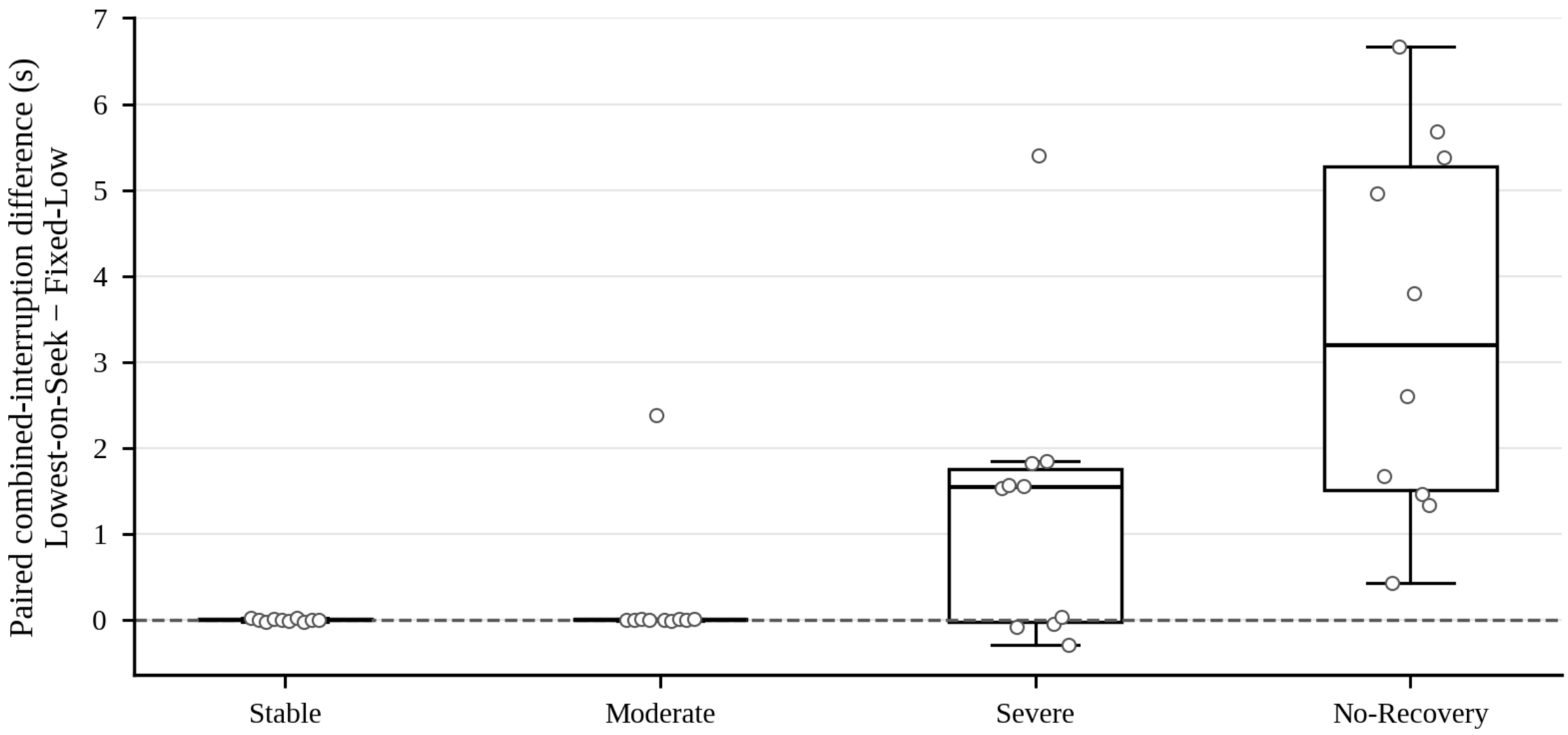

Figure 4B. Recovery persistence (RQ2): paired Lowest-on-Seek-minus-Fixed-Low combined-interruption differences.

Together, RQ1 and RQ2 distinguish the restart from the period that follows it. A low-cost first request helps playback resume, while keeping a conservative ceiling during buffer rebuilding can reduce later interruption. Brief stalls still occurred under Severe degradation.

## 5.3. RQ3: Historical-State Awareness in the Primary Campaign

RQ3 examines whether the inherited bandwidth estimate changes the recovery decision beyond what a fixed L0 ceiling achieves. Fixed-Low is the direct control, with Estimator-Reset included for context.

In the primary profiles, applying $\gamma$=0.15 to the inherited estimate produced an L0 ceiling for IAR in every block. IAR and Fixed-Low therefore used the same effective restriction in these runs.

The results reflect this. The median paired Fixed-Low-minus-IAR resume-delay difference was -1.7 ms, and the combined-interruption difference was -2.65 ms. Neither differed after Holm correction; the combined-interruption comparison had $p_{Holm}$=1.0. There was only one discordant stall block between the two strategies.

Estimator-Reset was similarly close to IAR on the primary continuity outcomes. The median Estimator-Reset-minus-IAR resume-delay difference was 1.8 ms, while the combined-interruption difference was 0.75 ms. Again, neither comparison remained different after family-wise correction.

This is a useful negative result. The primary campaign demonstrates the effects of a conservative first request and persistent recovery, but it cannot by itself establish graded sensitivity to inherited bandwidth state, because the naturally produced state happens to place IAR at the same L0 ceiling as Fixed-Low.

For this reason, historical-state sensitivity is tested causally in RQ5 by varying $B_h$ while holding the network profile fixed. Figure 4C shows the corresponding primary-campaign contrasts involving IAR.

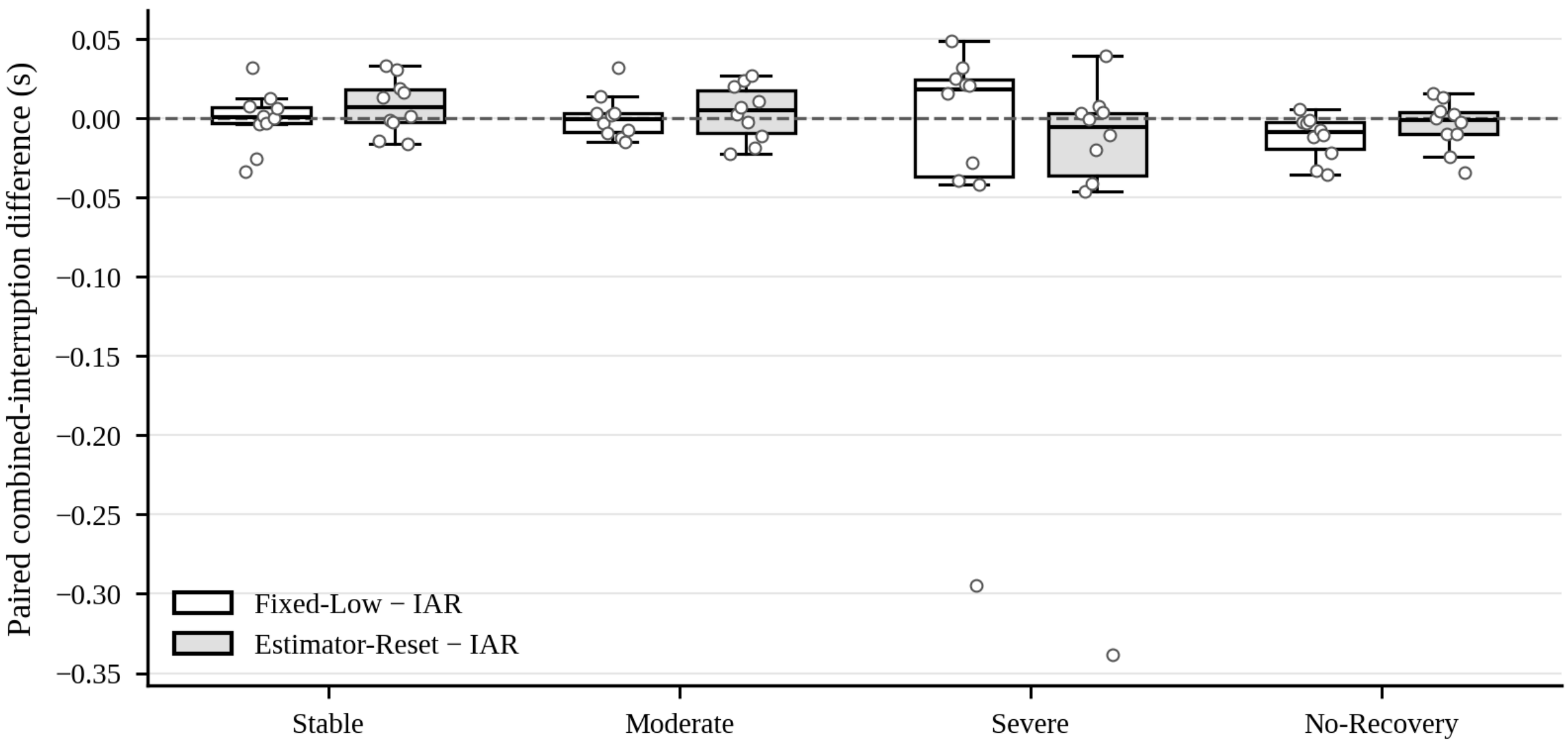

Figure 4C. Primary-profile historical-state context (RQ3): paired combined-interruption differences involving IAR.

## 5.4. RQ4: Presented Quality and Quality Recovery

RQ4 examines the quality cost associated with conservative recovery. Presented quality is evaluated using the bitrate of media actually displayed, rather than the nominal bitrate of successfully downloaded fragments. Presented-quality coverage was complete in every W20 and W40 observation window.

Figure 5A shows the time-weighted presented bitrate during W20. Relative to Lowest-on-Seek, IAR remains more conservative. For example, under Stable the median W20 bitrate was 3.466 Mbps for IAR and 4.048 Mbps for Lowest-on-Seek. Under Moderate the corresponding values were 2.187 and 2.591 Mbps; under Severe, 1.024 and 1.577 Mbps; and under No-Recovery, 1.170 and 1.972 Mbps.

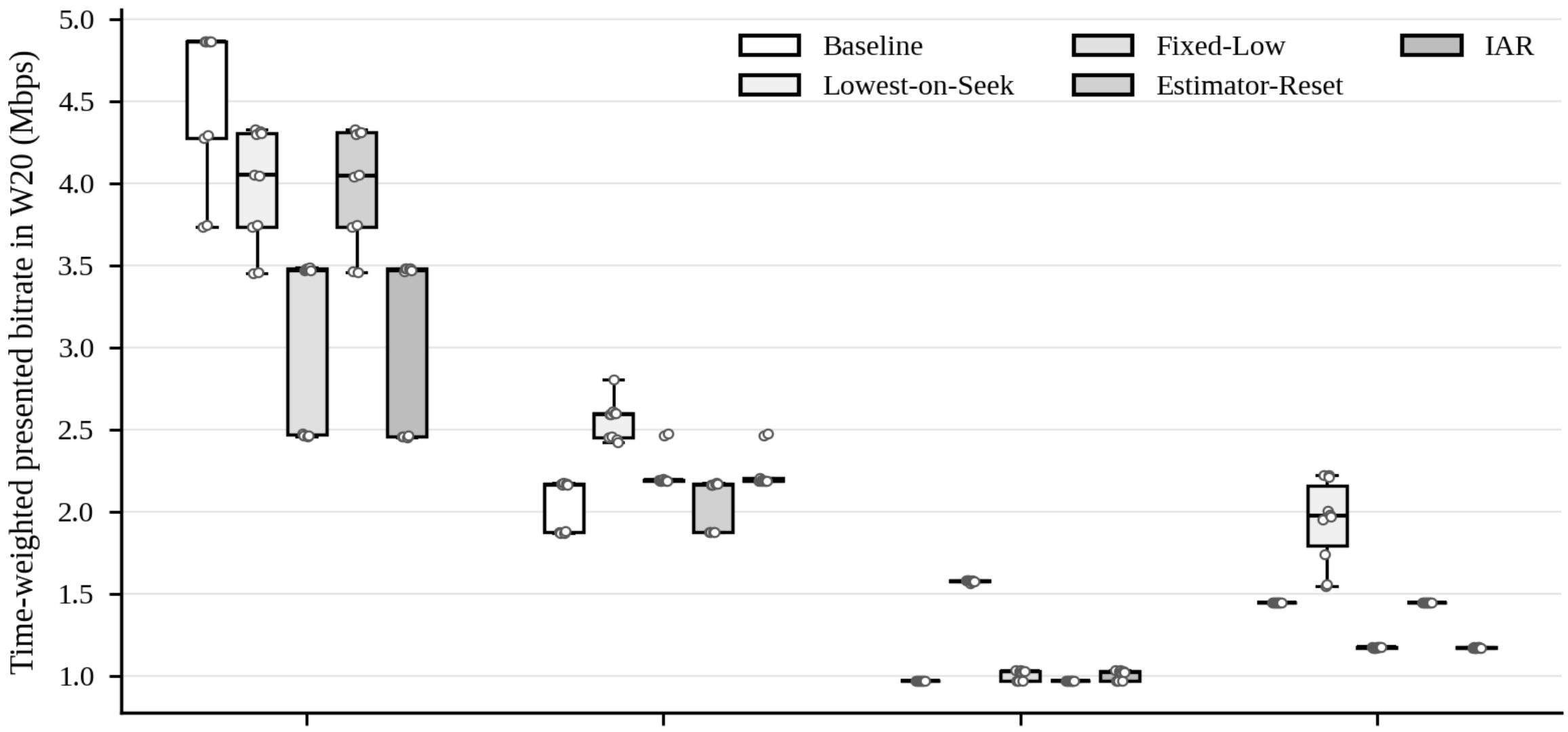

Figure 5A. Presented-quality cost during W20 after resumption (RQ4).

Across all 40 paired blocks, the median IAR-minus-Lowest W20 difference was -0.583 Mbps, with $p_{Holm}=7.28\times10^{-11}$. At W40, the corresponding paired difference was -0.216 Mbps, with $p_{Holm}=1.02\times10^{-6}$. The quality penalty therefore becomes smaller with time but remains visible over the extended window.

The conservative recovery also changes the path through the representation ladder. IAR produced one fewer W40 representation switch than Lowest-on-Seek in the median paired comparison, with $p_{Holm}=2.10\times10^{-9}$. The corresponding cumulative switch-magnitude difference was not distinguishable after Holm adjustment, indicating that the principal difference was in how often the level changed rather than in the total number of ladder steps traversed.

The comparison with Fixed-Low is again nearly exact: W20 bitrate, W40 bitrate, W40 switch count, switch magnitude, and quality-return outcome show no meaningful separation after correction. This is consistent with both strategies operating under the same L0 ceiling in the primary campaign.

Figure 5B reports return to the pre-seek representation. Under Stable, all five strategies returned to the pre-seek level within W20 and W40 in all ten blocks. Under Moderate, however, only Fixed-Low and IAR achieved the sustained-return condition in all 10 blocks; Baseline, Lowest-on-Seek, and Estimator-Reset did not regain the pre-seek level by W40 in any block. Under Severe and No-Recovery, no strategy regained the pre-seek level within W40.

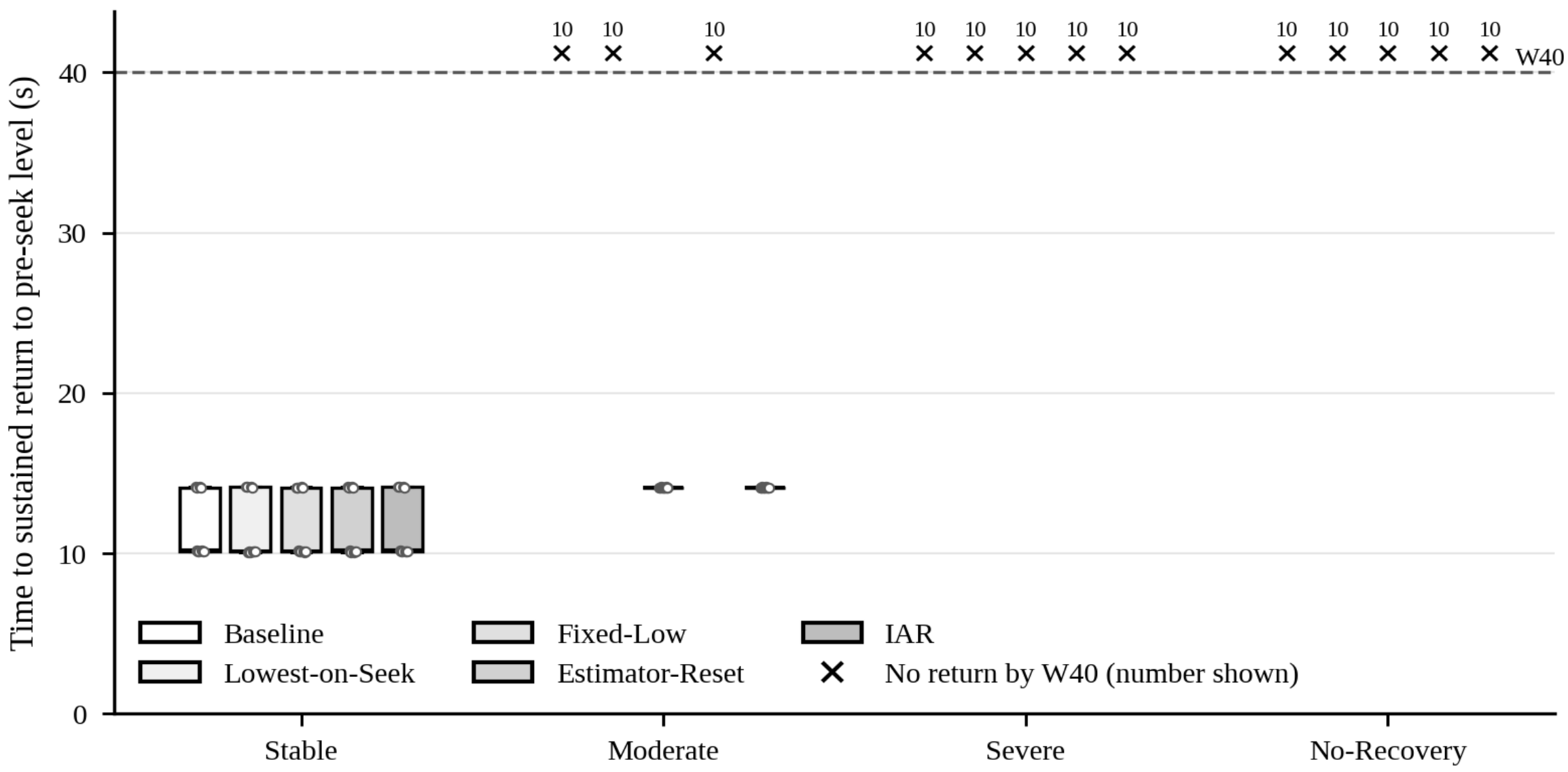

Figure 5B. Sustained return to the pre-seek representation within W40 (RQ4).

For IAR versus Baseline, Lowest-on-Seek, and Estimator-Reset, the W40 return outcome contained ten discordant blocks, all favoring IAR, and the exact McNemar comparison remained below 0.05 after Holm adjustment ($p_{Holm}$=0.0293). IAR and Fixed-Low had identical W40 return outcomes.

Return-delay values are interpreted only among runs that actually returned, because non-return by W40 is censored. For context, IAR's median sustained-return delay was about 10.15 s in Stable and 14.10 s in Moderate.

The Moderate-profile result may seem counterintuitive because the return measure requires four consecutive seconds of presented media at or above the pre-seek level, not merely a request or brief presentation at that level. The matched trajectory in Appendix A.5 illustrates a restricted-recovery path that later meets this criterion. The aggregate results, however, do not isolate which buffer or estimator transitions prevented the other strategies from meeting it within W40. We therefore report the measured return-pattern difference without assigning it to a single cause.

Compared with Lowest-on-Seek, IAR presented lower bitrate early after the seek and made fewer W40 level switches. Under Moderate degradation, it also met the sustained-return criterion in blocks where several less restricted strategies did not. These are differences in measured quality-return trajectories, not an established causal advantage of selecting lower bitrate.

Our measurements describe playback continuity and displayed representation, not subjective viewing quality. We did not conduct a user-perception study.

Appendix A.5 presents an illustrative trajectory from one matched Moderate-profile block. It clarifies the request, buffer, and displayed-quality sequence but is not additional inferential evidence.

## 5.5. RQ5: Controlled Sensitivity to Inherited Bandwidth State

The primary campaign did not vary IAR's selected ceiling. RQ5 examines that mechanism by controlling the scalar bandwidth prior at recovery entry.

The Moderate network profile was held fixed while the hls.js bandwidth estimator was reinitialized to scalar priors of 6, 10, 15, 20, 25, or 30 Mbps. Ten matched seek positions were evaluated at each prior, producing 60 runs. These values are controlled initializations rather than claims about naturally evolved EWMA histories.

The expected ceiling transitions occurred exactly. With γ=0.15, 6 and 10 Mbps mapped to L0, 15, 20, and 25 Mbps mapped to L1, and 30 Mbps mapped to L2. The first post-seek request and the first actually presented post-seek level matched the calculated ceiling in all 60 runs. Table 6 summarizes these controlled outcomes, while Figures 6A and 6B show the resume-delay response and the representation-ceiling transitions.

Table 6. Controlled historical-state sensitivity.

| Injected $B_h$ | $B_r$=0.15$B_h$ (Mbps) | Ceiling | Resume delay, s, median [IQR] | Recovery duration, s, median [IQR] | W20 bitrate, Mbps, median [IQR] | W40 bitrate, Mbps, median [IQR] |
|---|---|---|---|---|---|---|
| 6 Mbps | 0.90 | L0 | 0.876 [0.841, 0.912] | 2.600 [2.577, 2.674] | 2.027 [1.606, 2.031] | 2.569 [2.359, 2.571] |
| 10 Mbps | 1.50 | L0 | 0.876 [0.837, 0.911] | 2.607 [2.575, 2.673] | 2.026 [1.602, 2.031] | 2.568 [2.357, 2.571] |
| 15 Mbps | 2.25 | L1 | 1.514 [1.478, 1.563] | 5.823 [5.760, 5.865] | 1.788 [1.788, 1.791] | 2.450 [2.449, 2.451] |
| 20 Mbps | 3.00 | L1 | 1.504 [1.486, 1.564] | 5.819 [5.776, 5.853] | 1.791 [1.789, 1.791] | 2.451 [2.451, 2.451] |
| 25 Mbps | 3.75 | L1 | 1.516 [1.478, 1.569] | 5.812 [5.769, 5.863] | 1.789 [1.788, 1.791] | 2.450 [2.450, 2.451] |
| 30 Mbps | 4.50 | L2 | 2.650 [2.588, 2.722] | 7.022 [6.938, 7.098] | 2.524 [2.236, 2.526] | 2.817 [2.673, 2.818] |

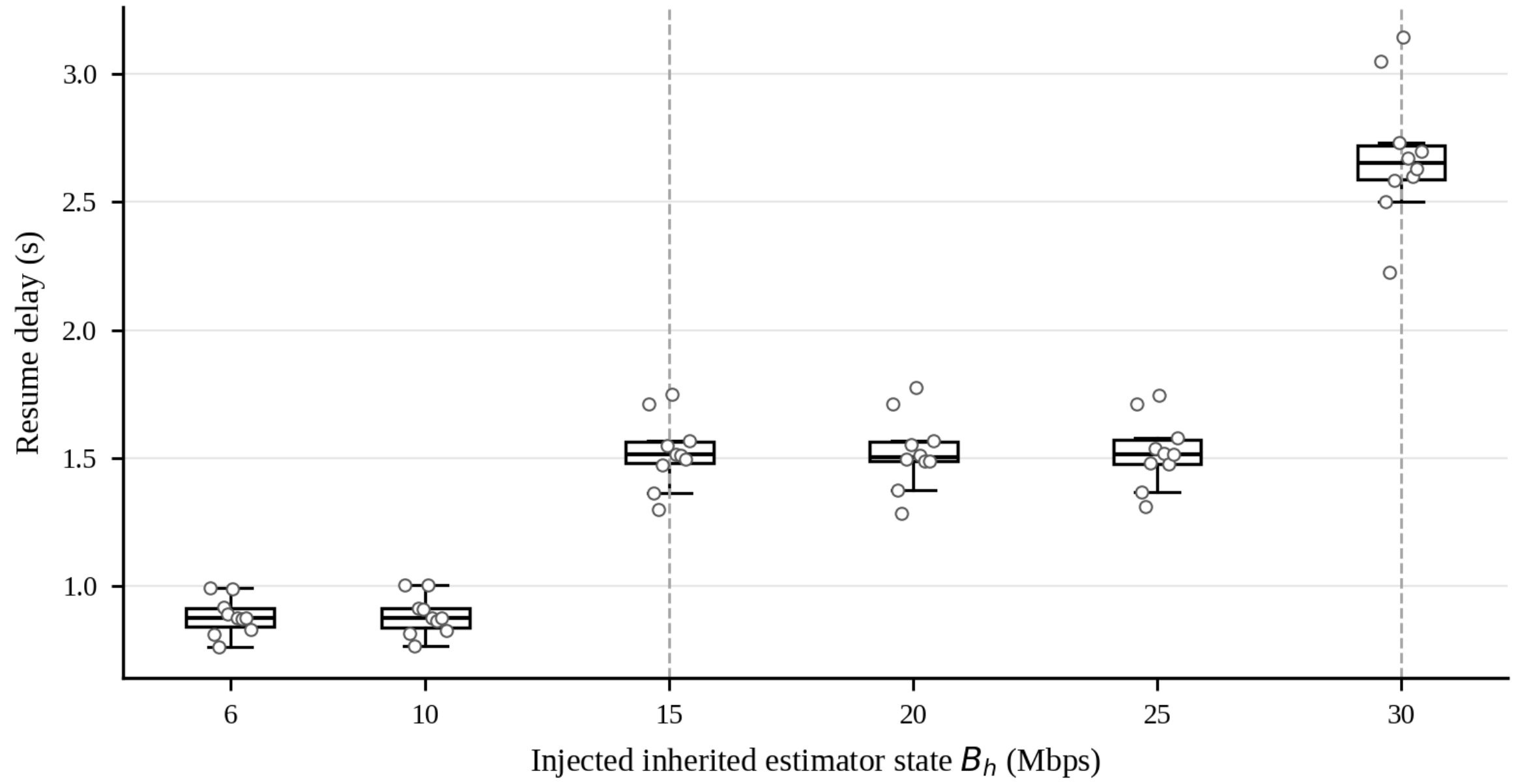

Figure 6A. Controlled historical-state sensitivity (RQ5): resume delay across inherited estimator states under the fixed Moderate profile.

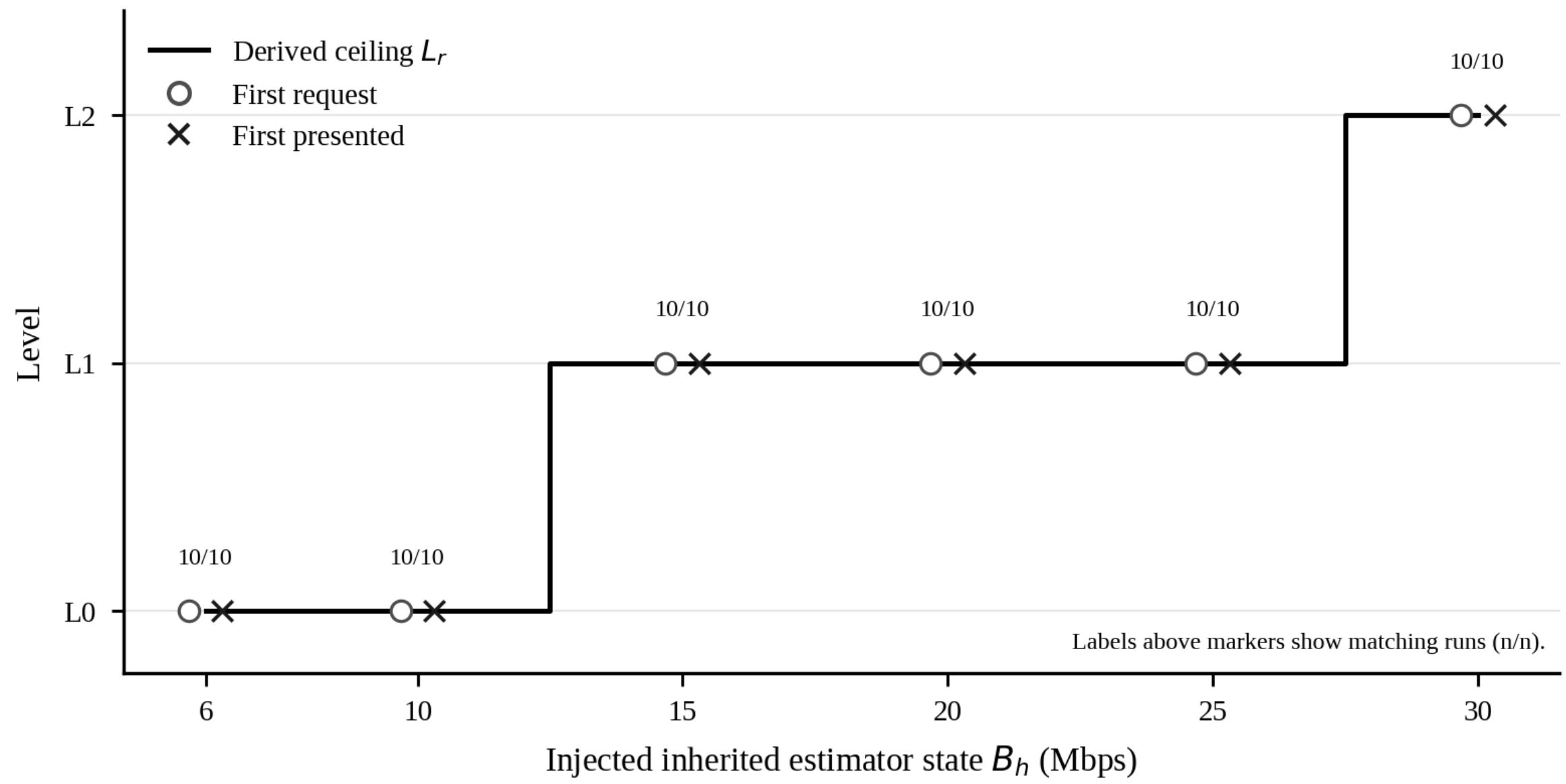

Figure 6B. Discrete representation-ceiling response to inherited estimator state (RQ5).

States that produce the same ceiling also produced similar resume times. For the two L0 states, 6 and 10 Mbps, the median paired difference was approximately 1 ms. Within L1, the 15-versus-20 and 20-versus-25 Mbps differences were about 2–3 ms. The main changes occurred when the prior crossed a representation threshold.

The first threshold crossing, from $B_h$=10 to 15 Mbps, changes the ceiling from L0 to L1. Resume delay increased by a median 640.0 ms, while recovery duration increased by 3192.4 ms. Both comparisons had rank-biserial magnitude 1.0 and remained below 0.05 after Holm correction ($p_{Holm}$=0.015625).

The second crossing, from 25 to 30 Mbps, changes the ceiling from L1 to L2. Resume delay increased by 1128.2 ms, and recovery duration increased by 1202.7 ms. At the same time, W20 presented bitrate increased by 0.735 Mbps and W40 bitrate by 0.367 Mbps. All four comparisons had $p_{Holm}$=0.015625.

The L2 condition also introduced an additional presented-quality down/up transition: the player began at L2, moved to L1, and later returned to L2, whereas the L0 and L1 conditions required fewer transitions during the same controlled profile.

No positive post-resume stalls occurred in the 60 RQ5 runs, including the ten runs with $B_h$=30 Mbps. The L2 ceiling was therefore feasible under the tested Moderate profile, although it increased resume and recovery times relative to less permissive ceilings while allowing higher presented bitrate.

RQ5 confirms that IAR responds to the scalar prior rather than always imposing an L0 ceiling. The bitrate response was not monotonic: the L1 plateau had lower median W20 and W40 presented bitrates than the L0 plateau, while the highest values occurred at L2 (Table 6). Because hls.js can select levels below the ceiling and continues adapting after recovery, a higher ceiling does not guarantee higher time-weighted presented bitrate. Resume and recovery times increased at the two tested upward ceiling transitions. These controlled results do not establish that a history-derived ceiling is preferable to Fixed-Low or Estimator-Reset across other networks.

## 5.6. Practical Implications and Limitations

Across the five strategies, the first request determines much of the restart delay, while the persistence and level of the restriction influence later continuity and displayed quality. IAR treats the unbuffered seek as a temporary recovery state while leaving steady-state hls.js adaptation in place.

These measurements come from CDP-based network emulation, one HLS presentation and four-level ladder, one browser environment, and hls.js 1.7.1. The experiments used controlled forward seeks to unbuffered locations. Different network traces, media ladders, segment durations, seek patterns, or ABR implementations may yield different magnitudes of effect.

RQ5 resets the hls.js estimator to controlled scalar priors under the Moderate profile; it does not reconstruct naturally evolved fast/slow EWMA histories. The 60 stall-free runs establish how the ceiling responds to these priors, not a failure threshold or a superiority claim over Fixed-Low in the primary campaign.

The primary audit identified eight negative STALL_END durations of less than 0.7 ms in magnitude from a mixed-clock instrumentation path. They were clamped to zero; the adverse ±0.7-ms sensitivity check did not change RQ2 stall classifications or its exact Wilcoxon p-values. The timing path was corrected before RQ5, although no RQ5 run stalled. Appendix A.3 and A.6 give the detailed checks and threats to validity.

The chosen $\gamma$=0.15 discount and 8-s exit threshold were fixed before the final campaigns and were not shown to be optimal in general. Previous work relates buffering to viewer engagement [26], but we did not measure engagement, subjective QoE, or computation overhead. Our measurements should be read as technical playback results. The fuller validity assessment appears in Appendix A.6.

# 6. Conclusion

At an unbuffered seek destination, the ABR controller has little forward-buffer protection but may still rely on bandwidth information from earlier transfers. IAR discounts that estimate to set a temporary representation ceiling and restores the previous cap after 8 s of forward buffer has been rebuilt.

In the 200-run paired evaluation, selecting a less expensive first post-seek request shortened resume delay relative to standard hls.js by a paired median of about 2 s. Keeping a conservative ceiling in place during buffer rebuilding reduced later interruption under stronger degradation. The benefit came with lower early presented bitrate and a different quality-recovery trajectory.

Across the primary network profiles, IAR and Fixed-Low behaved similarly because the discount mapped every inherited estimate to L0. The 200-run campaign therefore did not show a separate benefit from using history to choose the ceiling. In the 60 controlled RQ5 runs, changing the scalar prior moved IAR's ceiling across the predicted L0/L1/L2 thresholds. Resume and recovery times increased at the two upward transitions, while presented bitrate varied non-monotonically and was highest at L2. None of those runs stalled.

These findings motivate treating inherited bandwidth information differently during the short period after an unbuffered seek. They do not establish a universally suitable discount or buffer threshold, or a benefit in subjective quality of experience. Future work should examine fresh post-seek measurements, realistic network traces, other representation ladders and ABR players, more varied seek patterns, and direct viewer feedback.

# Appendix A. Reproducibility and Auxiliary Validation

## A.1. Effective hls.js ABR Configuration

The final player instantiated hls.js 1.7.1 with only debug=false explicitly overridden. Each accepted run recorded hls.userConfig and the complete effective hls.config in HLS_CONFIG_SNAPSHOT before source loading. Table A1 reports the ABR-related values most relevant to this study. Estimator-Reset therefore reset to the configured default estimate of 0.5 Mbps.

*Table A1. Key effective hls.js 1.7.1 configuration values recorded in the final runs.*

| Parameter | Effective value | Relevance |
|---|---|---|
| abrEwmaFastVoD | 3 s | Fast EWMA half-life for VoD bandwidth estimation |
| abrEwmaSlowVoD | 9 s | Slow EWMA half-life for VoD bandwidth estimation |
| abrEwmaDefaultEstimate | 500,000 bit/s (0.5 Mbps) | Default estimate used by Estimator-Reset |
| abrBandWidthFactor | 0.95 | Bandwidth safety factor for same/down selection |
| abrBandWidthUpFactor | 0.70 | Bandwidth safety factor for upward selection |
| abrMaxWithRealBitrate | false | Real measured bitrate is not used as the ABR max-bitrate basis |
| maxStarvationDelay | 4 s | Starvation constraint in ABR selection |
| maxLoadingDelay | 4 s | Maximum loading-delay parameter |
| maxBufferLength | 30 s | Configured forward buffer target/limit used by hls.js |
| capLevelToPlayerSize | false | No player-size cap on the available representation set |

## A.2. Recovery-Exit Ablation

The recovery-exit ablation used 24 runs: Moderate and Severe profiles × four seek targets (122, 266, 410, and 554 s) × three candidate rules. Table A2 pools the eight matched runs available for each rule. Buffer-only and combined were effectively indistinguishable in the tested conditions, whereas fragment-only released recovery much earlier without reducing stall incidence.

*Table A2. Quantitative summary of the recovery-exit ablation (8 matched runs per rule).*

| Exit rule | N | Median resume (ms) | Stall-positive runs | Median stall duration (ms) | Median recovery (ms) | Median exit buffer (s) |
|---|---|---|---|---|---|---|
| Fragment-only | 8 | 1465.4 | 2/8 | 0.0 | 2938.5 | 4.605 |
| Buffer-only | 8 | 1459.2 | 2/8 | 0.0 | 5342.4 | 8.274 |
| Combined | 8 | 1451.6 | 2/8 | 0.0 | 5328.5 | 8.292 |

Profile-specific checks led to the same interpretation. Under Moderate, all three rules were stall-free (0/4), while median recovery duration was 1.792 s for fragment-only, 2.646 s for buffer-only, and 2.643 s for combined. Under Severe, each rule had 2/4 stall-positive runs; median stall duration was 158.4, 163.2, and 163.0 ms, respectively, while median recovery duration was 4.234, 8.233, and 8.284 s. These results do not show that fragment-only is inferior to buffer-only. They show that adding the fragment-count condition to the 8-s buffer criterion did not provide observable additional protection in these matched runs. Buffer-only was retained because forward-buffer occupancy directly measures the playback protection that recovery is meant to restore, whereas a fixed fragment count is only an indirect proxy whose meaning depends on segment duration and concurrent playback consumption. The ablation therefore supports removing the fragment-count conjunct under the tested configuration, not a universal equivalence claim.

## A.3. Stall-Timestamp Perturbation Sensitivity

The primary campaign contained eight negative STALL_END artifacts whose magnitude was at most 0.7 ms. To test whether a clock offset of that observed magnitude could drive RQ2, an offline worst-direction perturbation was applied to every positive Lowest-on-Seek and Fixed-Low stall duration: one strategy was shifted by +0.7 ms while the other was shifted by -0.7 ms, and then the directions were reversed. No stall/non-stall classification changed. The exact Wilcoxon p-values for the two central RQ2 continuous outcomes were identical in both adverse perturbation directions. Table A3 summarizes the resulting sensitivity check.

*Table A3. Offline sensitivity of the RQ2 inference to the worst observed 0.7-ms timestamp offset.*

| RQ2 outcome | Original raw p | Adverse ±0.7-ms raw p | Reported Holm-adjusted result |
|---|---|---|---|
| Stall-duration difference | $2.6703 \times 10^{-5}$ | $2.6703 \times 10^{-5}$ | $1.07 \times 10^{-4}$ |
| Combined-interruption difference | $9.4117 \times 10^{-4}$ | $9.4117 \times 10^{-4}$ | 0.00282 |
| Stall occurrence | Discordant-pair classification unchanged | Unchanged | McNemar conclusion unchanged |

## A.4. Media-Artifact Scope

The frozen experiment artifacts verify the properties reported in Section 4.2: H.264 video, AAC 48-kHz stereo audio, 30 frames/s, 159 aligned media fragments per representation, fragments 0–157 of 4 s plus a final ~2.6-s fragment, and the exact BANDWIDTH/AVERAGE-BANDWIDTH ladder. The archived run-level snapshots do not retain the original encoder/packager command, GOP/keyframe interval, container-format declaration, or playlist independent-segment tag. Those details are therefore not reconstructed from memory, and bit-exact regeneration of the encoded presentation from the manuscript alone remains a reproducibility limitation.

## A.5. Illustrative Matched-Run Trajectory

Figures A1A–A1C illustrate one matched Moderate-profile block (s01, seek target 122 s) using the frozen primary logs. This example is descriptive, not additional inferential evidence. Baseline begins with an L3 request before later presenting L0 and climbing gradually; IAR begins under an L0 ceiling, rebuilds more than 8 s of forward buffer, releases the cap after about 2.63 s, and later satisfies the sustained quality-return condition.

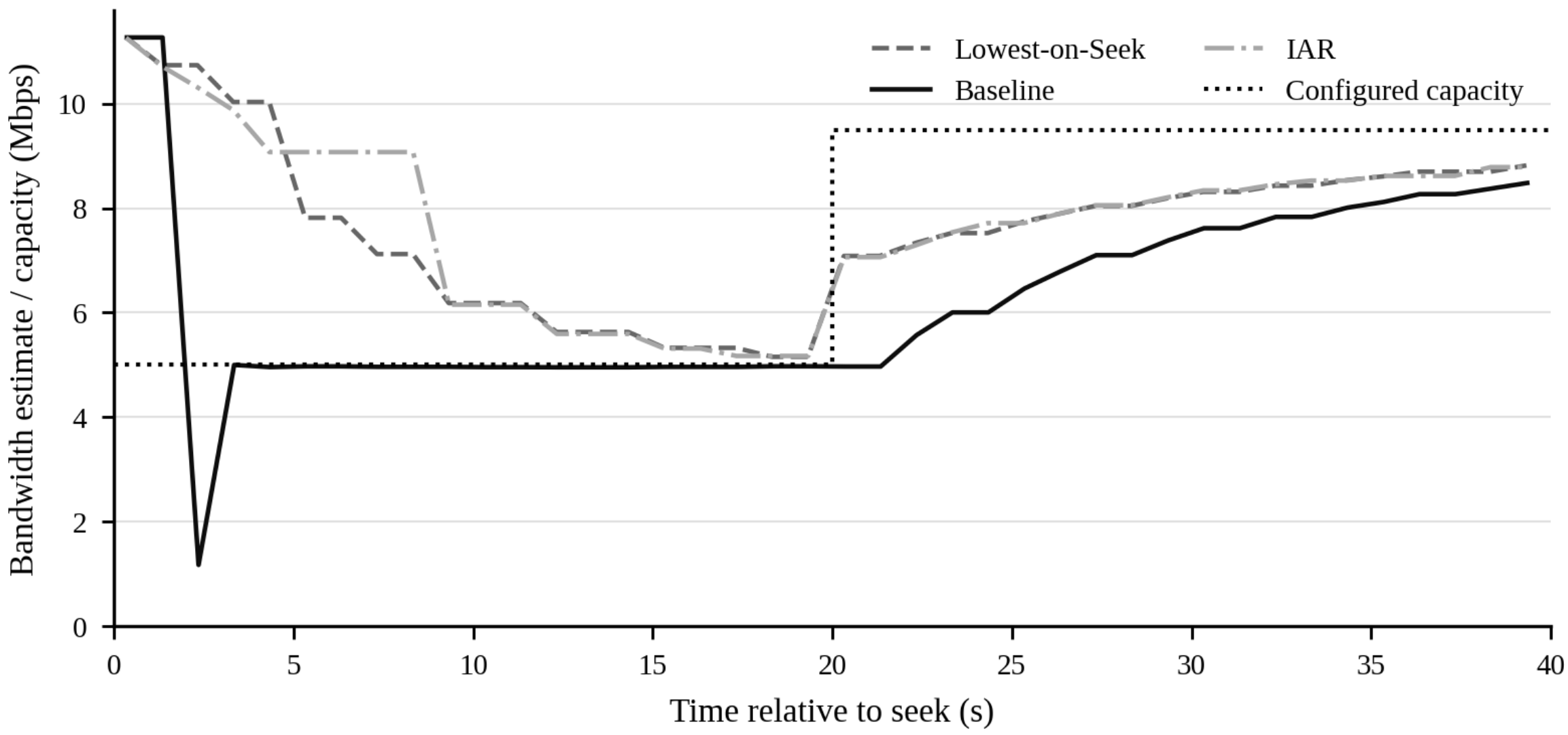

*Figure A1A. Illustrative actual-run trajectory: bandwidth estimate and configured capacity for Moderate block s01 (122-s seek).*

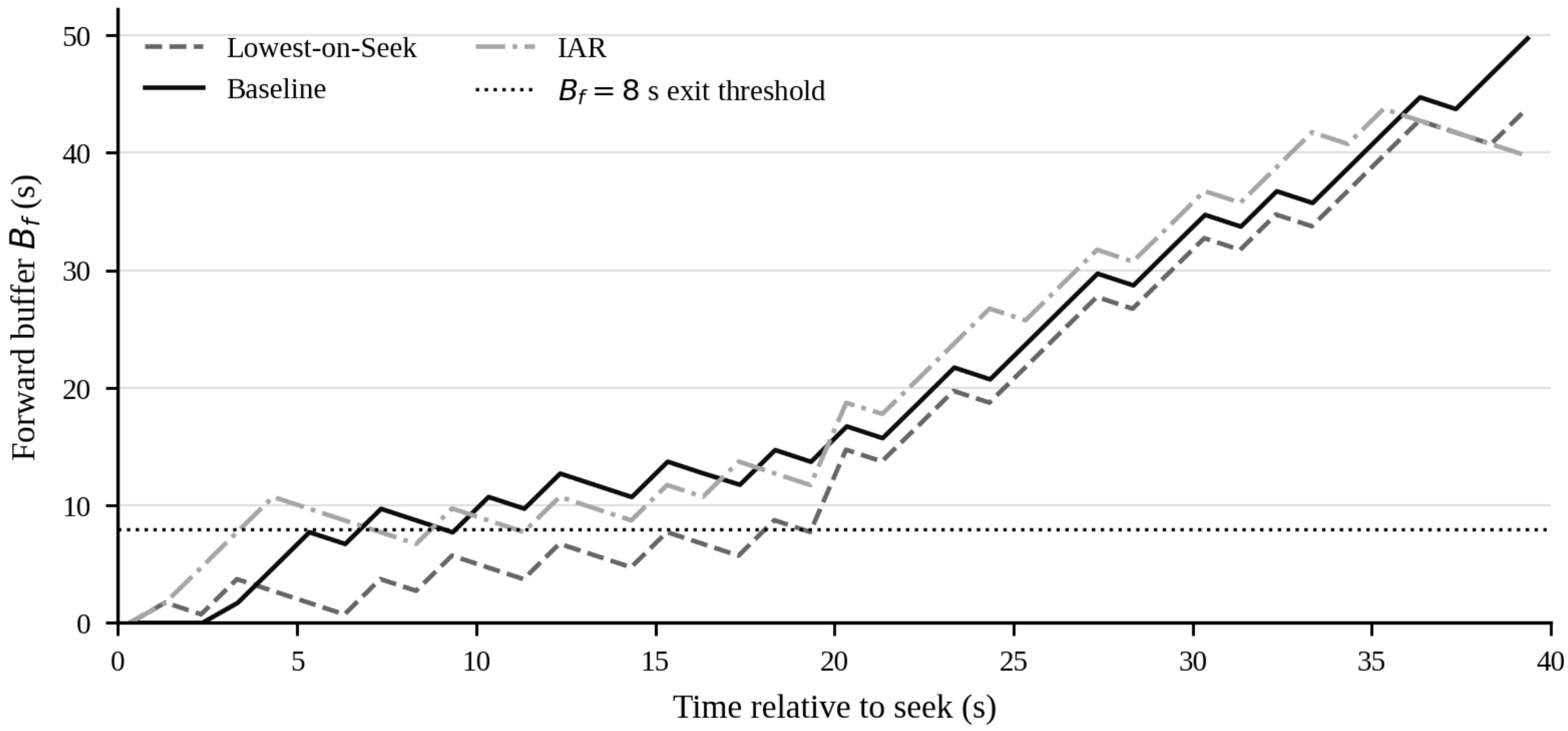

*Figure A1B. Illustrative actual-run trajectory: forward-buffer rebuilding in the same matched block.*

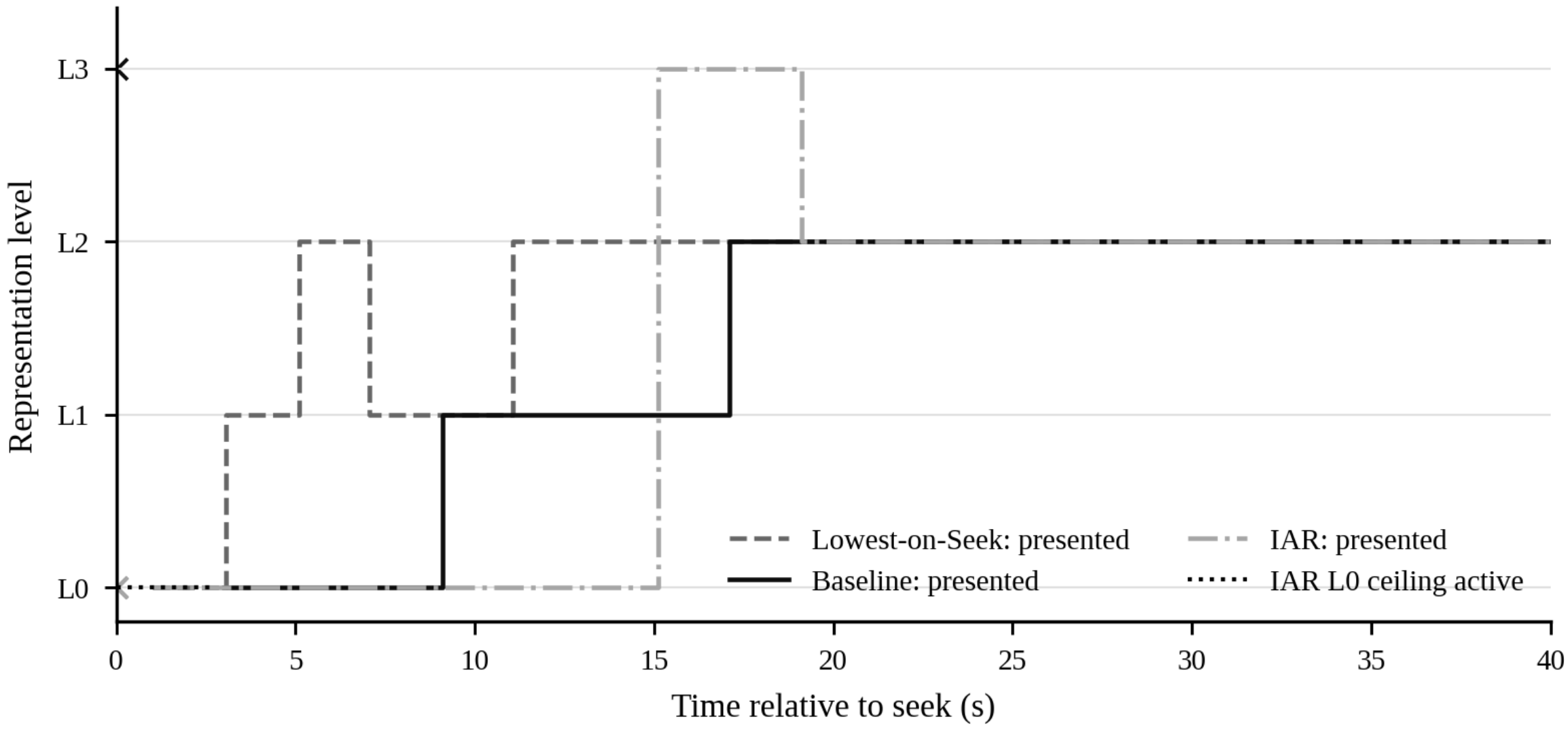

*Figure A1C. Illustrative actual-run trajectory: first request and presented-quality evolution in the same matched block.*

## A.6. Detailed Limitations and Threats to Validity

### A.6.1. Study Limitations

The controlled setup makes post-seek strategies directly comparable, but limits how far the numerical results can be generalized.

First, network conditions were imposed through browser-level Chrome DevTools Protocol emulation. The four primary profiles represent controlled changes in available capacity rather than complete models of real networks. Factors such as packet loss, short-term burstiness, competing cross traffic, wireless interference, mobility, and cellular handovers were not varied independently. Real network traces may therefore produce adaptation behavior that is not captured by the step changes used here.

Second, all experiments used one video presentation, one four-level HLS ladder, approximately 4-s fragments, H.264 video, AAC audio, and normal-speed playback. Representation spacing is particularly relevant to IAR because the continuous recovery budget $B_r$ is converted into a discrete representation ceiling. A denser or more widely spaced ladder could change the points at which the ceiling moves from one level to another. Segment duration can likewise affect both first-fragment delivery time and the rate at which forward buffer is rebuilt.

Third, the evaluation uses hls.js 1.7.1 and the ABR behavior of that implementation. IAR is deliberately implemented as a layer around the existing controller, so some observed behavior, including estimator evolution, emergency downswitches, and adaptation under a temporary cap, is specific to hls.js. The design principle could be implemented in other players, but the present results do not establish that the same numerical parameters or outcomes would transfer unchanged.

The interaction space is also restricted. Ten different seek destinations were evaluated, reducing dependence on one media position, but all final interactions are controlled forward seeks to unbuffered locations and all targets use approximately the same 2-s offset within their corresponding ~4-s segments. The study therefore varies content location but does not separately evaluate sensitivity to intra-segment seek position, backward seeks, repeated scrubbing, rapid successive seeks, or partially buffered destinations.

The final IAR parameters should not be interpreted as globally optimal. The value $\gamma$=0.15 was selected during development and frozen before the final campaigns. The 8-s forward-buffer threshold was also fixed for the final evaluation. The exit-rule ablation showed that, under the tested conditions, adding a two-fresh-fragment requirement to the 8-s buffer threshold did not provide observable additional protection. This finding supports the simpler final rule for the present configuration, but it does not establish that fragment count is redundant for every segment duration, representation ladder, or player.

RQ5 examines six controlled $B_h$ values, all under the same Moderate network trajectory. It identifies how the recovery ceiling changes with that scalar input, but does not test inherited state and post-seek capacity jointly. No stalls occurred in these 60 runs; that observation does not establish safety for more severe conditions or higher ceilings in other profiles.

We did not measure CPU use, memory consumption, or energy demand. The implementation adds limited control state, but its computational cost was not evaluated.

We did not conduct a subjective QoE study. The reported resume delays, stalls, switching, and presented bitrates describe technical playback behavior, not necessarily viewer preference. In particular, combined interruption weights a second spent waiting to resume the same as a second of stalling after resumption. The experiments cannot determine whether viewers would prefer one recovery trajectory over another.

## A.6.2. Threats to Validity

The main threats are discussed in terms of internal, construct, external, and conclusion validity, following the empirical-software-engineering framework already used in this manuscript [27].

### A.6.2.1. Internal Validity

Internal validity concerns whether differences attributed to the evaluated strategy may instead have been caused by other variation in the experimental environment.

Several controls were used to reduce this risk. All final runs used the same machine, browser version, hls.js version, media presentation, and encoding ladder. Browser caching was disabled, and network profiles were applied programmatically. Playwright automated browser startup, playback, seek execution, and experiment sequencing, reducing variation associated with manual execution.

The final campaign also used a frozen execution plan rather than collecting all observations of one strategy consecutively. Conditions were interleaved across the campaign, while the same profile/seek-position combinations were reused across strategies to preserve paired comparisons.

More importantly, the runner did not issue the experimental seek until the requested network condition had been applied successfully and the experiment had received positive acknowledgements for estimator-state injection, where applicable, and seek arming. This reduces ambiguity about whether a run began under the intended state.

The audit of the 200 primary runs found no missing conditions, duplicate strategy observations within blocks, seek-target mismatches, or fatal hls.js errors. All runs contained the required experimental seek, first post-seek request, first presented frame, and W20/W40 instrumentation.

The experiment also explicitly distinguishes controlled unbuffered seeks from other playback-position changes. Four 0.1-s position adjustments generated internally by hls.js during stalls were recorded in the primary campaign, but the logger identified them as unarmed seeks and excluded them from the experimental-seek path. This is preferable to the earlier displacement-threshold approach because the final trigger reflects the state actually under study: loss of useful buffered media at the destination.

One instrumentation issue remained in the primary campaign. Eight STALL_END records contained impossible durations between -0.7 and -0.5 ms because stall start and end were sampled from slightly different browser timing sources. They were clamped to zero and not counted as stalls. All other positive stalls were at least 55 ms. A worst-observed-offset sensitivity check (±0.7 ms applied adversarially to positive Lowest-on-Seek and Fixed-Low stall durations) left stall classifications and the RQ2 Wilcoxon p-values unchanged, so the reported RQ2 inference is not sensitive to an offset of the observed magnitude. This does not establish sub-millisecond timing accuracy, and the primary campaign was not rerun solely for this logger correction.

The logger was corrected before RQ5 so that stall start and end used the same performance.now() clock. RQ5 contained no negative stall durations, but no RQ5 run stalled, so the corrected STALL_END branch was not behaviorally exercised in that campaign. The correction therefore removes the identified mixed-clock path, while the offline perturbation analysis, not RQ5, supports robustness of the primary RQ2 conclusions to the observed sub-millisecond offset scale.

Another internal-validity consideration is that RQ5 deliberately reinitializes the hls.js bandwidth estimator through the bandwidthEstimate setter. In hls.js 1.7.1 this reset creates a new estimator initialized with the requested scalar prior; it does not preserve the fast/slow EWMA weights of a naturally evolved history. The intervention is therefore interpreted causally and narrowly: the network trajectory is fixed while the scalar value available to IAR at recovery entry changes. Conclusions from RQ5 concern sensitivity to that controlled prior, not the prevalence or full internal structure of naturally occurring estimator states.

Finally, the experiments were designed, implemented, executed, and analyzed within the same study rather than through a blinded or independently replicated process. Independent replication would provide additional confidence in the implementation and analysis.

### A.6.2.2. Construct Validity

Construct validity concerns whether the measurements used in the study represent the concepts the paper intends to evaluate.

Resume delay is defined from the controlled seek to the first valid advancing post-seek frame observed through `requestVideoFrameCallback()`. This is a stronger representation of actual playback resumption than relying only on an HTML media event, because the endpoint corresponds to a frame that was actually presented.

Post-resume stall duration measures interruption after that first resume. Nevertheless, stall detection remains an instrumentation-based operational definition and should not be equated automatically with perceived rebuffering severity.

Combined interruption, defined in Eq. (13), is useful for identifying strategies that merely move interruption from before playback resumes to afterward. However, it assumes equal technical weight for the two forms of waiting. A user may perceive a 1-s delayed resume differently from a 1-s stall after playback has already restarted. $T_i$ is therefore a continuity summary, not a QoE model.

We calculate presented quality for W20 and W40 from media actually displayed, rather than the nominal bitrate of downloaded fragments. This avoids assigning quality credit to a representation that was downloaded but never displayed. Still, the AVERAGE-BANDWIDTH value associated with each representation remains a nominal technical descriptor. It does not measure perceptual quality directly and does not account for content-dependent visual complexity.

The sustained quality-return metric also depends on an operational definition: the pre-seek level must be regained and maintained for 4 s of presented media. This avoids treating a momentary quality transition as recovery, but the 4-s value is a study definition rather than a perceptually validated threshold.

W20 and W40 are wall-clock windows beginning at the first resumed presented frame. Stalls therefore reduce the amount of presented media that can occur within a given wall-clock interval. This is intentional because the measure describes what is actually presented during the first 20 or 40 s experienced after resumption, but it differs from an analysis based on a fixed amount of media time.

For RQ5, the injected estimator state is also a construct that must be interpreted carefully. It is a reset-created scalar prior for the hls.js bandwidth estimator, not a reconstruction of a complete natural EWMA history, and it is not equivalent to configuring physical network bandwidth. The RQ5 results therefore establish sensitivity to the value supplied at recovery entry rather than the frequency or full trajectory of such states in deployment.

### A.6.2.3. External Validity

External validity concerns the extent to which the observed results can be generalized beyond the evaluated environment.

The study evaluates one ABR implementation, one browser family, one machine, one media asset, one representation ladder, one approximate segment duration, and four deterministic primary network profiles. The precise resume delays, stall durations, recovery times, and quality trajectories should therefore not be expected to reproduce unchanged in other systems.

Real network behavior may differ substantially from the controlled CDP profiles. Capacity can vary continuously rather than in discrete steps, and packet loss, queueing, jitter, congestion, and competing traffic can interact with throughput estimation in ways that are absent from the current testbed. Evaluation with public network traces, controlled physical links, and operational networks would strengthen external validity.

The media ladder also matters directly to RQ5. The observed L0/L1/L2 plateaus arise because the discounted bandwidth budget is mapped to the specific `BANDWIDTH` thresholds of the four-level experimental presentation. Another ladder would move these transition points.

The same applies to the 8-s exit threshold. With approximately 4-s fragments, 8 s corresponds to a particular amount of media protection. A player using shorter segments, low-latency parts, a different buffer policy, or a higher playback rate may rebuild and consume that protection differently.

Only forward unbuffered seeks were studied. Although ten seek destinations provide repeated evidence across the media presentation, the results do not directly address backward seeking, partially buffered destinations, repeated user scrubbing, or sequences of interactions before the previous recovery has completed.

RQ5 should also be generalized narrowly. It establishes that the implemented ceiling responds predictably to six controlled inherited states under the Moderate profile. It does not define a universal boundary between safe and unsafe historical estimates, because post-seek capacity was intentionally held fixed rather than varied jointly with $B_h$.

A broader evaluation would be needed to establish the best parameter settings. The general observation supported here is that the value of inherited bandwidth information can change when a seek removes forward-buffer protection.

### A.6.2.4. Conclusion Validity

Conclusion validity concerns whether the experimental design and statistical analysis support the conclusions drawn from the observed data.

The final primary campaign uses ten seek-position blocks for each of four profiles, yielding 40 paired blocks and 200 runs across the five strategies. RQ5 adds 60 controlled runs, with ten matched seek blocks for each of six estimator states. The inferential conclusions are therefore based on the paired block structure of the 200-run primary campaign and the matched 60-run RQ5 study.

Even so, the effective sample size for a within-profile strategy comparison is still ten paired blocks. This is sufficient to reveal large and consistent effects in the controlled setting, but it remains limited for making broad population-level claims about streaming behavior.

The statistical analysis preserves the paired experimental structure. Continuous paired outcomes use exact Wilcoxon signed-rank tests, binary outcomes use exact McNemar tests, and Holm adjustment is applied within predefined research-question families. The analysis also reports paired median differences and rank-biserial effects rather than relying only on significance tests.

Several conclusions depend more on the structure of the results than on a single $p$-value. For example, RQ5's main mechanistic evidence is the exact ceiling mapping across all 60 runs and the near-identical behavior of estimator values that fall within the same representation-ceiling interval. The threshold-crossing tests support this pattern, but the mechanism is also directly observable in the logged state transitions.

Multiple testing remains a concern because RQ4 includes several quality outcomes and strategy contrasts. Holm correction reduces the probability of treating isolated comparisons as significant merely by chance, but corrected non-significance should not be interpreted as proof that two strategies are equivalent. This is particularly relevant to IAR and Fixed-Low: their near-identical behavior in the primary campaign is consistent with their common L0 ceiling, but the study does not claim formal statistical equivalence.

Censoring also limits inference for quality-return time. Runs that do not regain the pre-seek representation by W40 do not have an observed return time. For that reason, raw return delay is reported descriptively, while inferential comparison uses the paired binary outcome `return-within-W40` rather than assigning censored runs an artificial value of 40 s.

The exit-rule ablation also has a limited scope. Buffer-only and combined exit behavior were effectively indistinguishable in the evaluated matched conditions, which justified simplifying the final algorithm. The result should not be interpreted as proof of equivalence across all network states, segment durations, and ladders.

Finally, none of the 60 controlled RQ5 runs stalled, so this experiment does not identify an estimator prior that causes stalling under the tested Moderate profile. It does show that a higher inherited prior can lead to a more permissive ceiling, longer resume and recovery times, and additional quality adaptation. The fixed discount therefore does not guarantee an optimal recovery ceiling, but the tested L2 condition cannot be described as unsafe based on these runs.

Overall, the evidence supports several specific conclusions within the evaluated environment: conservative first-fragment selection strongly affects resume delay; extending recovery beyond the first request can reduce later interruption; persistent recovery imposes a temporary quality cost; and IAR's ceiling responds predictably to inherited estimator state. The experiments do not establish that $\gamma$=0.15 or $B_f$=8 s is universally optimal, that IAR always outperforms standard ABR, or that the observed technical trade-offs necessarily correspond to improved subjective QoE.